\documentclass[aps,prd,twocolumn,showpacs,amsmath,nofootinbib]{revtex4-1}

\usepackage{color}
\newcommand{\beqa}{\begin{eqnarray}}
\newcommand{\eeqa}{\end{eqnarray}}
\usepackage{float}
\usepackage{epsfig}
\usepackage{graphicx}
\usepackage{latexsym}
\usepackage{amsmath}
\usepackage{mathrsfs}
\usepackage{dcolumn}
\usepackage{bm}%
\usepackage{multirow}

\begin{document}

\title{New cosmological constraints on the evolution of dark matter energy density}

\author{Yupeng Yang}\email{ypyang@aliyun.com}
\author{Xinyi Dai}
\author{Yicheng Wang}
\affiliation{$^1$School of Physics and Physical Engineering, Qufu Normal University, Qufu, Shandong, 273165, China}

\begin{abstract}
We constrain the evolution of dark matter energy density over time, specifically focusing on deviation from the standard model 
represented by the equation $\rho_{m}\propto(1+z)^{3-\varepsilon}$, where $\varepsilon$ is a constant parameter. 
Utilizing a diverse array of observational datasets, including baryon acoustic oscillations (BAO) data 
from the first release of the Dark Energy Spectroscopic Instrument (DESI), 
distance priors derived from cosmic microwave background (CMB) observations by the Planck satellite, Hubble rate data obtained through 
the cosmic chronometers (CC) method, type Ia supernova (SNIa) data from the Panthon sample, 
and the data from the redshift space distortion (RSD) measurements ($f\sigma_8$),
we derive stringent constraints on the deviation parameter. We find that for the model under consideration, the deviation parameter 
is constrained to be $\varepsilon  = -0.0073^{+0.0029}_{-0.0033}$, indicating a deviation of approximately $2.4\sigma$ from the 
scenario where dark matter and vacuum dark energy do not interact. 
When compared with previous studies and alternative analyses, our findings provide corroborative evidence for an interaction between 
dark matter and vacuum dark energy, particularly in light of the release of BAO data from DESI.

\end{abstract}

\maketitle

\section{Introduction}
Modern astronomical observations and research indicate that the majority of the Universe consists of dark matter and dark energy, 
whose natures remain unknown to us, accounting for 26\% and 69\% respectively~\cite{2020A&A...641A...6P}. 
In the standard cosmological model, known as the $\Lambda$CDM (Lambda Cold Dark Matter) universe, 
dark energy is considered as the so-called cosmological constant term $\Lambda$, characterized by a specific equation of state 
$w_{\Lambda}=p_{\Lambda}/\rho_{\Lambda}=-1$\cite{Copeland:2006wr}. 
Despite the remarkable success of the $\Lambda$CDM model, it is not devoid of challenges\cite{Perivolaropoulos:2021jda}. According to quantum field theory, 
if dark energy were composed of the vacuum, there would be a discrepancy of approximately 120 orders of magnitude between theoretical calculations 
and observed values. Furthermore, recent issues such as the Hubble tension have posed additional challenges to the standard $\Lambda$CDM model of 
the Universe~\cite{Hu:2023jqc,DiValentino:2021izs}. 
To address these challenges, numerous dark energy models have been introduced, with dynamical dark energy models being the most extensively 
researched~\cite{Copeland:2006wr,Wang:2016lxa,Elizalde:2021kmo,Khurshudyan:2024gpn,Khurshudyan:2023cim,Kumar:2025etf}. 
A key aspect of these models is that the equation of state parameter of dark energy is no longer a constant but evolve over time, i.e., $w=w(z)$. 
Additionally, it has been proposed that an interaction exists between dark energy and dark matter, see, e.g., Ref.~\cite{Wang:2016lxa} for a review. 

Dark energy interacts with dark matter through numerous mechanisms. One such mechanism involves the assumption 
that the cosmological constant is not static but rather time-varying, namely the vacuum decay model, 
denoted as $\Lambda=\Lambda(t)$, see, e.g., Refs.~\cite{Overduin:1998zv,PhysRevD.46.2404,Arbab:1999hv,Shapiro:2000dz,
Ray:2004nq,Espana-Bonet:2003qjh,Shapiro:2003ui,Gomez-Valent:2014rxa,Sola:2015wwa,Sola:2016jky,SolaPeracaula:2016qlq,SolaPeracaula:2017esw,Sola:2017znb,
Yang:2025vnm,PhysRevD462404,RGVishwakarma_2000,Arbab:1999hv,Kaeonikhom:2022ahf}. 
The variation of the cosmological constant term in the vacuum decay model can take different forms. 
Consequently, these variations lead to the evolution of dark matter energy density over time, which diverges from the standard form of $\rho_{dm}\propto(1+z)^3$. 
Note that this scenario was first proposed by the authors of~\cite{Shapiro:2003ui} and 
subsequently tested using the SNIa data in~\cite{Espana-Bonet:2003qjh}.
The exact manner of deviation is determined by the specific form of interaction. 
Alternative forms of dark matter that evolve differently from standard models can be found in Refs.~\cite{1984MNRAS.211..277D,Doroshkevich,cmun,10.1093/mnras/239.3.923,Doroshkevich:1984gw,Wu:2025wyk,Wu:2024faw}. 
In contrast to this notion, the authors of~\cite{Wang:2004cp} suggested that it is not necessary to specify the exact form of interaction. 
Specifically, if the vacuum undergoes decay into dark matter, the resultant energy density of dark matter evolves over time with a modified power-law relation, 
given by $\rho_{dm}\propto(1+z)^{3-\varepsilon}$, where $\varepsilon$ is a positive constant. Recently, 
the authors of~\cite{SantanaJunior:2024cug} investigated this scenario and 
constrained the parameter $\varepsilon$ utilizing measurements of the gas fraction in galaxy clusters. In this work, 
we revise the constraints on the deviation parameter $\varepsilon$ by employing a diverse range of data sources. 
These include baryon acoustic oscillations (BAO) data from the Dark Energy Spectroscopic Instrument (DESI), 
distance priors derived from cosmic microwave background (CMB) observations by the Planck satellite, 
Hubble rate measurements obtained through the cosmic chronometers (CC) method, type Ia supernova (SNIa) data, 
and the data from the redshift space distortion (RSD) measurements.  
Our analysis reveals stronger constraints on $\varepsilon$, suggesting that the evolution of the energy density of dark matter closely adheres 
to the standard case within the currently considered model. 

This paper is organized as follows. In Sec. II we first provide a brief overview of the basic equations related to the deviation parameter. 
Subsequently, we describe the datasets and methods employed in our analysis. Finally, we derive new constraints on the deviation parameter. 
The conclusions are given in Sec. III.


\section{The basic equations of model}

In this section, we briefly review the main contents of used model in this work, and one can refer to, e.g., 
Refs.~\cite{Koussour:2024nhw,Oztas:2018zvi,Vishwakarma:2000gp,Overduin:1998zv,Azri:2014apa,Azri:2012fpi,Szydlowski:2015bwa,Bruni:2021msx,SolaPeracaula:2023swx,Bora:2021uxq,Bora:2021iww,Holanda:2019sod} for more details. 
As mentioned above, the deviation of dark matter energy density from its standard case can be realized within the framework of vacuum decay cosmology. 
The vacuum decay model has been proposed by previous works, see, e.g., Refs~\cite{Overduin:1998zv} for a review. 
In the classical cosmology model, i.e. $\Lambda$CDM model, $\Lambda$ does not evolve with time. 
The key point of the vacuum decay model is that vacuum energy density changes over time $\Lambda(t)$.  
Within the framework of the standard cosmological model under the Friedmann-Robertson-Walker (FRW) metric, the continuity equation 
can be formulated as follows~\cite{SolaPeracaula:2017esw,Brito:2024bhh,Feng:2019jqa},

\beqa
&&\dot{\rho}_{dm}+3H\rho_{dm}=Q \\
&&\dot{\rho}_{\rm \Lambda}=-Q
\eeqa
where $\rho_{dm}$ and $\rho_{\Lambda}$ are the energy density of dark matter and vacuum. 
$Q$ denotes the interaction between dark matter and vacuum dark energy, and $Q=0$ for $\Lambda$CDM model. 
The special form of the deviation depends on the form the interaction between the vacuum decay and dark matter~\cite{Wang:2016lxa}. 
In this study, we consider the case where 

\beqa
Q=\varepsilon H\rho_{dm}
\label{eq:interaction}
\eeqa

Note that we here assume 
that there is no interaction between baryonic matter and the vacuum. Consequently, the evolution of baryonic matter adheres to the standard scenario, 
and the same applies to radiation~\cite{SolaPeracaula:2017esw}.

Rather than adopting a specific form for $Q$, the authors of~\cite{Wang:2004cp} proposed that the interaction could be 
described by the extent to which the dark matter energy density deviates from its standard evolution. 
This deviation is mathematically represented as $\rho_{dm}=\rho_{m0}(1+z)^{3-\varepsilon}$, where $\varepsilon$ is a constant. 
A positive value of $\varepsilon$ indicates a continuous decay of vacuum energy into dark matter, while a negative value represents the opposite process. 
For flat $\Lambda$CDM cosmology, the expansion rate of the Universe, $H(z)$, can be expressed as follows~\cite{Shapiro:2003ui,Espana-Bonet:2003qjh,Wang:2004cp},

\beqa
H^{2}=H_{0}^{2}\Bigg[\frac{3\Omega_{dm,0}}{3-\varepsilon}(1+z)^{3-\varepsilon}+\Omega_{b}(1+z)^{3}+ \Omega_{\gamma}(1+z)^{4} \nonumber \\
+1-\frac{3\Omega_{dm,0}}{3-\varepsilon}-\Omega_{b}-\Omega_{\gamma}\Bigg]
\label{eq:hz}
\eeqa
with $\Omega_{i}\equiv 8\pi G\rho_{i}/(3H_{0}^{2})$ is the energy density parameter of different component at present. 

Note that although the authors of~\cite{Wang:2004cp} did not specify a precise expression for 
the interaction (i.e., a specific form for $Q$), 
it is still feasible to derive a general form of the interaction based on 
the evolutionary behavior of the dark matter energy density. For instance, the authors of~\cite{Wang:2004cp} proposed 
an interaction form given by $d{\Lambda}/da = {\varepsilon}/[24{\pi}G(1+{\varepsilon})]dR/da$, 
where $R$ represents the scalar curvature of spacetime. Additionally, the authors of~\cite{Barboza:2024gcd} 
identified an equivalent form of the interaction, expressed as $Q={\varepsilon}H\rho_{dm,0}(1+z)^{3-\varepsilon}$.

Since we will also utilize data from redshift space distortion measurements, which are associated with the evolution of 
density perturbations, the evolution of the matter perturbation $\delta_{m}$ for various forms of $Q$ has been extensively 
derived in prior research, e.g., Refs.~\cite{PhysRevD.90.023008,Sola:2016jky,Gomez-Valent:2018nib,SolaPeracaula:2017esw}. 
In this study, we employ the methodologies presented in~\cite{SolaPeracaula:2017esw}. 
Specifically, the evolution of $\delta_{m}$ as a function of the scale factor $a$ can be expressed as follows~\cite{SolaPeracaula:2017esw}: 

\beqa
\frac{d^{2}\delta_m}{da^2}+\frac{A(a)}{a}\frac{d\delta_m}{da}+\frac{B(a)}{a^{2}}\delta_m=0
\label{eq:delta_a}
\eeqa
where the function $A(a)$ and $B(a)$ are given by 

\beqa
A(a) = 3+\frac{a}{H(a)}\frac{dH(a)}{da}+\frac{Q/{\rho}_m}{H(a)}
\eeqa

\beqa
B(a) = -\frac{4\pi G\rho_{m}}{H^{2}(a)}+\frac{2Q/{\rho}_m}{H(a)}+\frac{a}{H(a)}\frac{d(Q/\rho_m)}{da}
\eeqa
here $\rho_{m}=\rho_{dm}+\rho_{b}$, and $H(a)$ is the Hubble rate given by Eq.~(\ref{eq:hz}). 
$Q$ is the interaction between 
dark matter and vacuum dark energy given by Eq.~(\ref{eq:interaction}). For the model under consideration herein, the values of $\delta_{m}(a)$ and 
its derivative $d\delta_{m}(a)/da$ at a specific scale factor $a$ or redshift $z=1-1/a$ can be determined by numerically 
solving the above equations. Note the evolution of $\delta_m$ presented in~\cite{Wang:2004cp} (specifically, Eq.~(19)) 
differs from that described in~\cite{SolaPeracaula:2017esw} (namely, Eqs.~(28),(29), and (30)). 
When compared to the latter set of equations, the former omits certain small quantities. 
These equations serve as the foundational framework for analyzing Redshift Space Distortion (RSD) data, 
and a detailed exposition can be found in the subsequent section, specifically~\ref{sec:rsd}.

\section{datasets and constraints on parameters }

\subsection{Baryon acoustic oscillation}

The Dark Energy Spectroscopic Instrument (DESI) has made public its first-year data release (DR1), 
encompassing six diverse categories of traces~\cite{DESI:2024mwx}. These include: 
the Bright Galaxy Sample (BGS,$z_{\rm eff}=0.30$), the Luminous Red Galaxy Sample (LRG, $z_{\rm eff}=0.51$ and 0.71), 
the Emission Line Galaxy Sample (ELG, $z_{\rm eff}=1.32$), 
the combined LRG and ELG Sample (LRG+ELG, $z_{\rm eff}=0.93$), 
the Quasar Sample (QSO, $z_{\rm eff}=1.49$) and the Lyman-$\alpha$ Forest Sample (Ly$\alpha$, $z_{\rm eff}=2.33$). 
The data released by the DESI collaboration is presented in the form of $D_{\rm M,H,V}/r_{d}$, where $r_{d}$ is the sound horizon at the drag epoch~\cite{DESI:2024mwx}. 
Within the framework of a homogeneous and isotropic cosmology, the transverse comoving distance $D_{\rm M}(z)$ can be expressed as follows
~\cite{2020A&A...641A...6P},

\beqa
D_{\rm M}(z) = \frac{c}{H_0}\int^{z}_{0}\frac{dz^{'}}{H(z^{'})/H_0},
\eeqa
where $c$ is the light speed, while $H(z)$ denotes the Hubble parameter, which is specified by Eqs.~(\ref{eq:hz}). 
The distance variable is defined as $D_{\rm H}(z)=c/H(z)$. Subsequently, the angle-averaged distance $D_{\rm V}$ can be expressed as 

\beqa
D_{\rm V}(z) = \left[zD_{\rm M}(z)^{2}D_{\rm H}(z)\right]^{\frac{1}{3}}
\eeqa

The $\chi^{2}$ for DESI BAO data is

\beqa
\chi^{2}_{\rm BAO}=\sum_{i}\Delta D_{i}^{T}{\rm Cov}^{-1}_{\rm BAO} \Delta D_{i},
\eeqa
where ${\rm Cov}_{\rm BAO}$ is non unit covariance matrix for the tracers of LGR,LGR+ELG, ELG and Ly$\alpha$ QSO, 
and a unit matrix for the tracers of BGS and QSO~\cite{Li:2024hrv}.  

\subsection{Cosmic microwave background}
In general, the angular power spectrum data of CMB have served as a means to constrain cosmological parameters~\cite{2020A&A...641A...6P}. 
However, for our analysis, we opt to use distance priors derived from the Planck 2018 data. These priors include the 
shift parameter $R$, the acoustic scale $l_{\rm A}$,and the baryon density $\Omega_{\rm b}h^{2}$~\cite{2020A&A...641A...6P,Chen:2018dbv}. 
The shift parameter and acoustic scale can be written as~\cite{2020A&A...641A...6P,Liu:2018kjv,Xu:2016grp,Feng_2011} 

\beqa
&&R=\frac{1+z_\star}{c}D_{\rm A}(z_{\star})\sqrt{\Omega_{\rm M}H^{2}_0} \\ 
&&l_{\rm A}=(1+z_{\star})\frac{\pi D_{\rm A}(z_{\star})}{r_{s}(z_\star)}
\eeqa
here $z_\star$ is the redshift at the epoch of photon decoupling and we adopt the approximate form of~\cite{Hu:1995en} 

\beqa
z_{\star} =1048[1+0.00124(\Omega_{b}h^{2})^{-0.738}][1+g_{1}(\Omega_{m}h^{2})^{g_{2}}]
\eeqa
with 

\beqa
g_{1}=\frac{0.0738(\Omega_{b}h^{2})^{-0.238}}{1+39.5(\Omega_{b}h^{2})^{0.763}},~ 
g_{2}=\frac{0.560}{1+21.1(\Omega_{b}h^{2})^{1.81}}
\eeqa

The comoving sound horizon $r_s$ can be expressed as follows~\cite{DESI:2024mwx} 

\beqa
r_{s}(z)=\frac{c}{H_{0}}\int^{1/(1+z)}_{0}\frac{da}{a^{2}H(a)\sqrt{3(1+\frac{3\Omega_{b}h^{2}}{4\Omega_{\gamma}h^{2}}a)}}
\eeqa
where $(\Omega_{\gamma}h^{2})^{-1}=42000(T_{\rm CMB}/2.7{\rm K})^{-4}$ with $T_{\rm CMB}=2.7255\rm K$. The $\chi^{2}$ for CMB data is

\beqa
\chi^{2}_{\rm CMB}=\Delta X^{T}{\rm Cov}^{-1}_{\rm CMB} \Delta X
\eeqa
where $\Delta X=X-X^{\rm obs}$ is a vector with $X^{\rm obs}=(R,l_{\rm A},\Omega_{b}h^{2})$, 
and we use the values of $X^{\rm obs}$ and covariance matrix ${\rm Cov}^{-1}_{\rm CMB}$ from Planck 2018~\cite{Zhai:2018vmm}.

\subsection{Hubble rate from cosmic chronometers}

The Hubble rate can be ascertained through the application of the cosmic chronometer (CC) method, 
as outlined in, e.g., the Refs.~\cite{Jimenez:2001gg,Moresco:2022phi,Jimenez:2023flo}. 
This technique primarily involves measuring the differential age evolution of the Universe, denoted as $dt$, 
over a specified redshift interval $dz$~\cite{Jimenez:2001gg}. In practice, 
this is achieved by conducting detailed studies of carefully chosen massive and passive galaxies. 
The Hubble rate, when determined using the CC method, can be expressed as follows~\cite{Jimenez:2001gg}

\beqa
H(z)=-\frac{1}{1+z}\frac{\Delta z}{\Delta t}
\eeqa

The $\chi^{2}$ for CC data is 

\beqa
\chi^{2}_{\rm CC}=\sum_{i}\frac{(H(z_i)-H_{\rm obs}(z_i))^{2}}{\sigma^{2}_i},
\eeqa
and for our constraints, we utilize 31 CC data points given in different Refs.
~\cite{Li:2019nux,2010JCAP...02..008S,2012JCAP...08..006M,Moresco:2016mzx,Moresco:2015cya,Ratsimbazafy:2017vga,2014RAA....14.1221Z,Singirikonda:2020ieg}.


\subsection{Type Ia supernova}
For type Ia supernova (SNIa) observations, a key parameter is the luminosity distance, denoted as $d_{L}(z)$. 
For a universe described by the flat $\Lambda$CDM model, the luminosity distance can be expressed as 
\beqa
d_{L}(z) = (1+z)\int^{z}_{0}\frac{dz^{'}}{H(z^{'})/H_0}
\eeqa

Additionally, the distance modules $\mu(z)$ is given by 

\beqa
\mu(z) = 5{\rm log}_{10}d_{L}(z)+25
\label{eq:sn}
\eeqa
 
The $\chi^{2}_{\rm SN}$ is formulated as follows, 

\beqa
\chi^{2}_{\rm SN}= \sum_{i} \frac{(\mu_{th}(z_{i})-\mu_{obs}(z_{i}))^{2}}{\sigma^{2}_{i}}   
\eeqa
where the theoretical values, $\mu_{th}(z_i)$, are computed using Eq.~(\ref{eq:sn}). 
$\mu_{obs}(z_i)$ denotes the observational data obtained from SNIa. 
For our constraints, we utilize the Pantheon sample~\footnote{https://github.com/dscolnic/Pantheon}, 
which comprises 1048 SNIa data spanning a redshift range of $0.01<z<2.3$~\cite{Pan-STARRS1:2017jku}.

\subsection{Redshift sapce distortion}\label{sec:rsd}

The redshift space distortion (RSD) data typically pertains to measurements $f\sigma_{8}(a)$, which are derived from 
various observations of the large scale structure~\cite{Song:2008qt,Davis:2010sw,Hudson_2012,Turnbull_2011,Samushia_2012,Blake_2012,Tojeiro_2012,Chuang_2013,
Beutler_2012,Blake_2013,S_nchez_2014,Chuang_2016,de_la_Torre_2013,Howlett_2015,Feix:2015dla,Okumura:2015lvp,Huterer:2016uyq,Pezzotta:2016gbo,BOSS:2016wmc}. 
Here, $f(a)$ represents the linear growth rate of matter perturbations $\delta_{m}(a)$, and it can be expressed as follows~\cite{Nesseris:2017vor,SolaPeracaula:2017esw}:

\beqa
f(a) \equiv \frac{a}{\delta_{m}} \frac{d\delta_m}{da}=\frac{d{\rm ln}\delta_m}{d{\rm ln}a}
\eeqa

$\sigma_{8}(a)$ is the root mean square mass fluctuation within a sphere of radius $8h^{-1}$ Mpc, which can be written 
as follows~\cite{Khatri:2024yfr,Nesseris:2017vor,SolaPeracaula:2017esw}:

\beqa
\sigma_{8}(a) =\sigma_{8}(a=1)\frac{\delta_{m}(a)}{\delta_{m}(a=1)}
\eeqa

Utilizing the relationship $a=1/(1+z)$, the weighted linear growth can be written as~\cite{Khatri:2024yfr}, 

\beqa
f\sigma_{8}(z) =-(1+z)\frac{\sigma_{8,z=0}}{\delta_{m,z=0}}\frac{d\delta_{m}}{dz}
\eeqa
here, the derivative $d\delta_{m}/dz$ can be determined by numerically solving Eq.~(\ref{eq:delta_a}). 
To solve this equation, we employ the initial condition specified in~\cite{daSilva:2020mvk}, where 
$\delta_{mi}$ is approximately $10^{-4}$ at $a_{i}\sim 10^{-3}$, and $d\delta_{mi}/da$ is set to $\sim \delta_{mi}/a_{i}$. 
This choice of initial conditions ensures that the matter perturbations remain within the linear regime.

In line with previous studies, e.g., Refs.~\cite{Khatri:2024yfr,Nesseris:2017vor,Sahlu:2023wvl,Kazantzidis:2018rnb}, 
we consider $\sigma_{8,z=0}$ as a free parameter in our analysis. 
Given that the $f\sigma_8$ data has been measured under the assumption of a fiducial $\Lambda$CDM cosmology, 
the corresponding values of $f\sigma_8$ in a true or alternative cosmology should be expressed as~\cite{Kazantzidis:2018rnb}:

 \beqa
f\sigma_{8}(z) =\frac{H(z)D_{A}(z)}{H_{\rm fid}(z)D_{A,{\rm fid}}(z)}f\sigma_{8,\rm fid}(z)
\eeqa

The $\chi^{2}$ for $f\sigma_{8}$ data is

\beqa
\chi^{2}_{f\sigma_{8}}=\sum_{i}\frac{(f\sigma_{8}(z_i)-f\sigma_{8,\rm obs}(z_i))^{2}}{\sigma^{2}_i},
\eeqa

For our analysis and constraints, we employ 22 data points as provided in, e.g., Ref.~\cite{Barboza:2024gcd}. 
It is important to note that when utilizing data from Wigglez and SDSS-IV, the correlations among the data points 
must be taken into account. The corresponding covariance matrices for these datasets can be found in, e.g., Refs.~\cite{Blake2012TheWD,10.1093/mnras/sty2845}. 
Note that the the methodology employed in this study diverges from that utilized in, for instance, Ref.~\cite{SolaPeracaula:2017esw}. 
The implications of these differences on the ultimate outcomes will be elucidated in the subsequent section.


\subsection{Constraints on the parameters}
\begin{figure}
\centering
\includegraphics[width=0.48\textwidth]{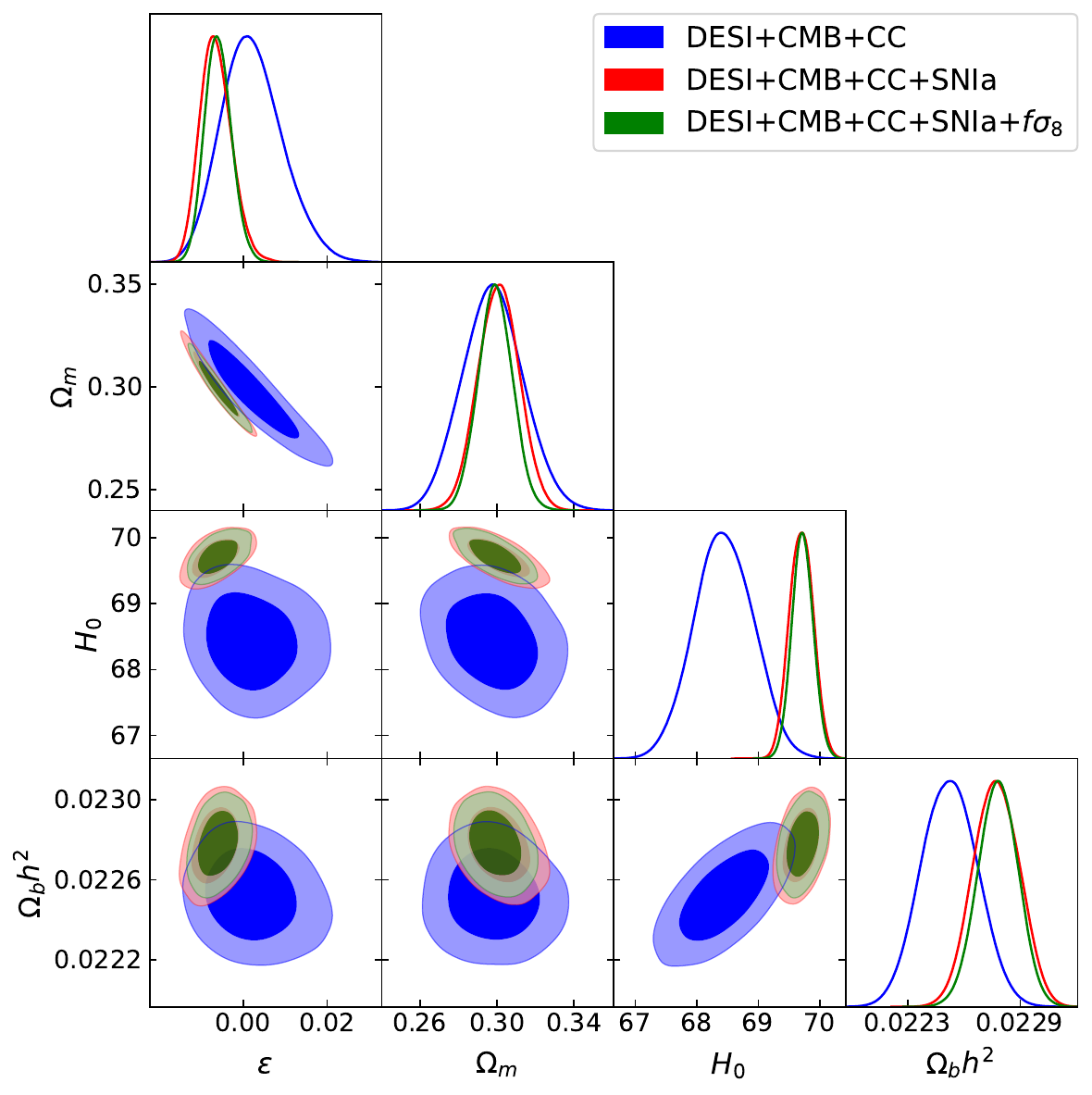}
\caption{The one-dimensional marginalized probability distribution and two-dimensional contour plots of parameters 
at the 68\% and 95\% confidence levels, 
derived from various observational data combinations: DESI+CMB+CC, DESI+CMB+CC+SNIa, and DESI+CMB+CC+SNIa+$f\sigma_8$. }
\label{fig:cons}
\end{figure}

For our purpose, we utilize the Markov Chain Monte Carlo (MCMC) method to determine the best fit values and the posterior distributions of 
parameters: \{$\varepsilon$, $\Omega_{m}$, 
$H_0$, $\Omega_{b}h^2$, $\sigma_8$\}. Given that we will incorporate data from DESI BAO, CMB, CC, SNIa, and $f\sigma_8$, the overall $\chi^{2}$ is computed as 

\beqa
\chi^{2}_{\rm total}=\chi^{2}_{\rm BAO} + \chi^{2}_{\rm CMB} + \chi^{2}_{\rm CC} + \chi^{2}_{\rm SN} + \chi^{2}_{ f\sigma_8}
\eeqa
The total likelihood function is $L\propto e^{-\chi^{2}_{\rm total}/2}$, and 
we use the publicly available $\mathtt{emcee}$~\cite{emcee} code to generate MCMC samples. 
We set uniform prior for the parameters as $\varepsilon \in (-1,1)$, $\Omega_{m}\in (0,1)$, $H_{0}\in (50,90)$, 
$\Omega_{b}h^{2} \in (0.001,0.1)$. For $\sigma_8$ we utilize a gaussian prior of $0.760 \pm{0.023}$~\cite{Heymans:2020gsg}. 
For analyzing the MCMC chains, we employ the $\mathtt{GetDist}$ public code~\cite{getdist}. analyze the MCMC chains. 
The best fit parameters, along with their corresponding $1\sigma$ uncertainties, are presented in Tab.~\ref{table:cons}. 
Fig.~\ref{fig:cons} displays the one-dimensional marginalized probability distributions and two-dimensional contour plots for the parameters, 
obtained from both the DESI+CMB+CC, DESI+CMB+CC+SNIa, and DESI+CMB+CC+SNIa+$f\sigma_8$ datasets. 

Our analysis reveals that the deviation parameter $\varepsilon$ exhibits a preference for a positive value and is constrained to 
$0.0023^{+0.0055}_{-0.0067}$ when utilizing the DESI+CMB+CC datasets. This suggests that there is no significant deviation 
from the standard evolution of dark matter energy density. 
However, when incorporating SNIa data into the analysis (i.e., using the DESI+CMB+CC+SNIa datasets), 
$\varepsilon$ shows a preference for a negative value, with constraints of 
$-0.0080^{+0.0033}_{-0.0041}$, indicating a deviation from the standard evolution at approximately $2.2\sigma$.  
Furthermore, a similar analysis with a slightly different constraint of 
$-0.0073^{+0.0029}_{-0.0033}$ for the DESI+CMB+CC+SNIa+$f\sigma_8$ datasets 
also suggests a deviation from the standard evolution, this time at approximately $2.4\sigma$ significance.

Compared to the constrained results obtained from the DESI+CMB+CC+SNIa datasets, the inclusion of $f\sigma_8$ data results in a slight shift of 
the central value of $\varepsilon$ towards larger value, specifically from -0.0080 to -0.0073. 
Additionally, this inclusion results in reduced uncertainties, a trend that is also evident in Fig.~\ref{fig:cons}. 
Additionally, the Hubble constant, $H_0=69.65\pm{0.17}$, lies between the values derived from the late-redshift and high-redshift universe. 
The value of $\sigma_{8,z=0}$ is constrained to $0.762\pm{0.018}$. It should be noted that, since we have utilized the shift parameters 
from CMB rather than the angular power spectrum data for our constraints, 
there are no specific constrains on $\sigma_8$ for the DESI+CMB+CC and DESI+CMB+CC+SNIa datasets. 

We also take into account the model selection criteria by employing the Akaike Information Criterion (AIC)
~\cite{Akaike1974A,1978AnSta...6..461S,Kass01061995}, 
which is mathematically defined as follows: 

 \beqa
{\rm AIC} =\chi^{2}_{\rm min} + \frac{2nN}{N-n-1}
\eeqa
where $n$ denotes the number of free parameters in the model, and $N$ represents the total number of data points utilized 
for the specific cosmology model under consideration. To evaluate the preference for a particular model over the benchmark $\Lambda$CDM 
model using the available datasets, we calculate the difference $\Delta {\rm AIC}$, define as 
$\Delta {\rm AIC} = {\rm AIC}_{\rm model} - {\rm AIC}_{\rm \Lambda CDM}$, with the $\Delta \rm AIC$ value of the $\Lambda$CDM model 
set to zero (${\Delta \rm AIC}_{\rm \Lambda CDM}=0$). The interpretation of $|\Delta \rm AIC|$ values in model comparison is 
as follows~\cite{SolaPeracaula:2017esw}: 
When $|\Delta \rm AIC|<2$, the two models are practically equivalent; 
when $2<|\Delta \rm AIC|<6$, the model with the larger AIC value has marginally less support compared to the model with the smaller AIC; 
when $6<|\Delta \rm AIC|<10$, the model with the larger AIC value is substantially less favored; 
finally,  when $|\Delta \rm AIC|>10$, the model with the larger AIC value is effectively unsupported by the data, 
and the evidence strongly favors the model with the smaller AIC. 

The values of $\Delta \rm AIC$ corresponding to various datasets are presented in Table~\ref{table:cons}. 
For the dataset combination of DESI, CMB, and CC, the calculated $\Delta \rm AIC$ value is $+2.36$. 
This result suggests that, although the data offer some support for the interacting model under investigation, 
the $\Lambda$CDM model remains the preferred choice due to its lower AIC value. 
In contrast, when analyzing the dataset combinations of DESI+CMB+CC+SNIa and DESI+CMB+CC+SNIa+$f\sigma_8$ 
the $\Delta \rm AIC$ values are found to be $-3.39$ and $-2.82$, respectively. 
These negative values unequivocally indicate a statistically stronger empirical preference for the interacting model 
currently under consideration relative to the $\Lambda$CDM model in these cases.


\begin{table*}[htp]
\caption{Constraints on the parameters \{$\varepsilon$, $\Omega_{m}$, 
$H_0$, $\Omega_{b}h^2$, $\sigma_8$\} with $1\sigma$ uncertainty, derived from different combinations of observational data 
for the same model currently being analyzed: DESI+CMB+CC, DESI+CMB+CC+SNIa, and DESI+CMB+CC+SNIa+$f\sigma_8$. 
Note that since we have utilized the distance
priors derived from Planck-2018 data rather than the angular power spectrum data for our constraints, 
there are no specific constrains on $\sigma_8$ for the first two datasets. 
The differences in Akaike Information Criterion values, denoted as $\Delta \rm AIC$ and calculated as 
$\Delta \rm AIC = AIC_{\rm model} - AIC_{\Lambda CDM}$, are presented to facilitate model comparison. 
A negative $\Delta \rm AIC$ indicates that the model under consideration provides a better fit to the observed datasets 
compared with the $\Lambda$CDM model.
}
\label{table:cons}
\begin{center}     
\begin{ruledtabular}  
\begin{tabular}{lccccccc}
Dataset & $\varepsilon$& $\Omega_{m}$ & $H_0$ &$\Omega_{b}h^{2}$ &$\sigma_8$ & $\Delta$AIC\\
\noalign{\smallskip}\hline\noalign{\smallskip}         

DESI+CMB+CC   &$0.0023^{+0.0055}_{-0.0067}$ &$0.297\pm{0.014}$  &$68.43\pm{0.43}$ &$0.02252\pm{0.00014}$ &$-$ & $+$2.36 \\
            
DESI+CMB+CC+SNIa  &$-0.0080^{+0.0033}_{-0.0041}$ &$0.3040\pm{0.011}$  &$69.63\pm{0.19}$ &$0.02282\pm{0.00012}$ &$-$ & $-$3.39  \\

DESI+CMB+CC+SNIa+$f\sigma_8$  &$-0.0073^{+0.0029}_{-0.0033}$ &$0.3016\pm{0.0090}$  &$69.65\pm{0.17}$ &$0.02282\pm{0.00011}$ &$0.762\pm{0.018}$ & $-$2.82 \\

\end{tabular}
\end{ruledtabular}
\end{center}
\end{table*}


For the sake of comparison, we have also included in Table~\ref{table:compare} the constraints on $\varepsilon$ 
that were documented in several previous studies which made use of varied datasets. The detailed explanations 
regarding these constraints are provided below. 

\begin{itemize}
\item[$\bullet$]In Ref.~\cite{SolaPeracaula:2017esw}, the authors conducted an investigation into three vacuum decay models, one of which (denoted as $Q_{dm}$) 
is pertinent to our model. In their study, they set $Q_{dm}=3\nu_{dm}H\rho_{dm}$. By employing data from SNIa (JLA), previous BAO measurements, 
CC (H($z$)), CMB data from Planck-2015, and information pertaining to large-scale structure formation ($f(z)\sigma_{8}(z)$), 
they determined that $\nu_{dm}$ is constrained to $0.00216\pm{0.00060}$. This constraint can be 
translated into a constraints on $\varepsilon$, yielding a value of $0.0065\pm{0.0018}$. This finding suggests a deviation of 
approximately $3.6\sigma$ from the scenario where there is no interaction between dark matter and the vacuum energy.

\item[$\bullet$]The authors of~\cite{Wagoner:2020wht} suggested that cluster edges can serve as a standard ruler and demonstrated that a 1.3\% 
measurement of the Hubble constant could be achieved using data from the DESI survey. Additionally, they generated a mocked dataset for the DESI survey. 
In contrast, the author of~\cite{Wang:2020hqq} combined these mocked DESI data with CMB (distance priors from Planck-2018), previous BAO, CC, SNIa (Pantheon), and 
mocked gravitational wave data from the Einstein Telescope to derive constraints on $\varepsilon$. The result, 
$-0.43^{+0.95}_{-1.30}$, indicates no significant interaction between dark matter and the vacuum. 

\item[$\bullet$]The authors of~\cite{Wang:2018ahw} obtained constraints by combining CMB data, specifically 
temperature and polarization data from Planck-2015, with previous BAO, CC, SNIa (Pantheon), 
and Planck-2015 lensing data (LS). The resulting value of $\varepsilon = -0.00029^{+0.00028}_{-0.00025}$ 
suggests a very slight deviation from the scenario where there is no interaction between dark matter and vacuum dark energy.

\item[$\bullet$]The authors of~\cite{SantanaJunior:2024cug} also 
derived constraints on $\varepsilon$ utilizing the measurements of gas mass fraction in galaxy clusters. For their purposes, they used the previous 
BAO, CC and SNIa data to reconstruct $D_{A}(z)$ and $D_{L}(z)$. Their analysis yielded constraints of $\varepsilon=0.191\pm{0.091}$. 
This constraint is less stringent compared to our results but suggests a deviation of 
approximately $ 2.1\sigma$ from the scenario where there is no interaction between dark matter and the vacuum energy.

\item[$\bullet$]Research conducted on the angular diameter distance of galaxy clusters, as presented in~\cite{Wang:2017fcr}, 
and measurements of HII galaxies, detailed in~\cite{Wang:2016pag}, have also provided robust constraints. 
Specifically, these studies yielded values of  $0.0019\pm{0.0089}$ and $0.010^{+0.005}_{-0.007}$ respectively. 
The latter result suggests a deviation of approximately $1.7\sigma$ from the scenario where there is no interaction 
between dark matter and vacuum dark energy.

\item[$\bullet$]The authors of~\cite{Kaeonikhom:2022ahf} obtained the constraints on the parameter $\varepsilon$ by utilizing the CMB data from Planck-2018, 
particularly the temperature and polarization data from Planck-2018, in conjunction with SNIa (Pantheon), previous BAO, and $f\sigma_8$ measurements 
from the SDSS database. 
The obtained result, $\varepsilon = 0.0016\pm{0.0014}$, indicates a slight deviation of approximately $1.1\sigma$ from the standard scenario of $Q=0$. 
   
\item[$\bullet$]The authors of~\cite{Li:2024qso} obtained the constraints on $\varepsilon$ using the CMB data, 
specifically temperature and polarization data from Planck-2018, with DESI BAO (DR1), and SNIa (DESY5) datasets. 
The resulting value of $\varepsilon = -0.0003\pm{0.0011}$ suggests no interaction between dark matter and vacuum dark energy.

\item[$\bullet$]Recently, the authors of~\cite{Pan:2025qwy} conducted a analysis to constrain the parameter $\varepsilon$ 
using various datasets. These datasets included the CMB shift parameter from Placnk-2018, CC, SNIa from Pantheon+, Union3, and DESY5, 
as well as the DESI BAO (DR2). 
Their results indicated a preference for non-zero interaction at a significance level of up to $2\sigma$. 
Specifically, for the combined dataset of CMB+DESI+CC+Pan+(S)~\footnote{Here, Pan+(S) denotes the Pantheon+ dataset calibrated using SH0ES.}, the value of $\varepsilon$ was constrained to $0.0257\pm{0.0034}$. 
This finding suggests an interaction between dark matter and vacuum dark energy at a significance exceeding $3\sigma$. 
For the datasets of CMB+DESI+CC+DESY5, the value of $\varepsilon$ was found to be $0.0073\pm{0.0035}$, 
indicating an interaction between dark matter and vacuum dark energy at a significance of $2.09\sigma$. 

However, it is important to note that for certain datasets, as shown in their paper, there was no strong evidence of interaction between dark matter and vacuum dark energy. 
For instance, when using the CMB+DESI BAO+CC+Pan+(Union3) datasets, they found that $\varepsilon$ was constrained to $-0.0008\pm{0.0041}$ ($-0.0002\pm{0.0041}$), 
which is essentially consistent with the scenario of no interaction ($Q=0$).

\end{itemize}

From the aforementioned results, it is clear that the most of the constraints employed the dataset comprising BAO, CMB, CC, and SNIa. 
Generally, different data and methodologies tend to produce distinct optimal fitting values and corresponding errors for $\varepsilon$. 
The constraints placed by the CMB, SNIa, and previous BAO dataset on $\varepsilon$ did not exhibit significant deviations from the standard 
scenario where there is no interaction between dark matter and vacuum dark energy. However, the inclusion of LSS ($f\sigma_8$) and DESI BAO datasets 
imposes more stringent constraints on $\varepsilon$, resulting in a noticeable deviation from the non-interaction scenario. 
The degree of this deviation depends on the specific datasets and methodologies employed.

Furthermore, it is important to highlight the utilization of CMB and LSS ($f\sigma_8$) data. For CMB data, a common practice involves employing both temperature 
and polarization data, as seen in, e.g., Refs.~\cite{SolaPeracaula:2017esw,Wang:2018ahw,Kaeonikhom:2022ahf,Li:2024qso}. 
Alternatively, CMB distance priors, which effectively capture the essential information from CMB temperature and polarization data, 
are also frequently utilized, as demonstrated in, e.g., Refs~\cite{Wagoner:2020wht,Pan:2025qwy,Liu:2018kjv,Xu:2016grp}. 
This latter method is comparatively more efficient and can reduce the time required for fitting compared to the former approach. 
Nevertheless, it is noteworthy that the method utilizing CMB temperature and polarization data yields smaller errors in the 
constraint of $\varepsilon$ compared to the distance priors method. However, the former method (utilizing the full set of CMB data) 
also tend to converge towards smaller 
optimal values of $\varepsilon$~\footnote{Note that the value of $\varepsilon$ referred to here represents its absolute value. 
Since it depends on the dataset, the optimal fitting value of $\varepsilon$ may be either greater than 0 or less than 0.}, 
resulting in minimal changes to the final deviation. For the parameter $\varepsilon$, the incorporation of LSS ($f\sigma_8$) information significantly 
improves the final constraint results. Particularly when combined with DESI BAO data, CMB, and SNIa datasets, 
the deviation of $\varepsilon$ from zero becomes more evident, suggesting an interaction between the dark matter and vacuum dark energy. 
This finding is consistent with the indications from the second data release of DESI showing a deviation from the $\Lambda$CDM model~\cite{DESI:2025zgx}.

For utilizing LSS ($f\sigma_8$) data, there are general two methods. The first approach, as employed in Refs.~\cite{SolaPeracaula:2017esw,Kaeonikhom:2022ahf,SolaPeracaula:2021gxi}, for instance, 
involves directly calculating $\sigma_8(z)$ through the integration of the linear matter power spectrum within a smooth window in Fourier space, 
which incorporates the spectral index $n_s$ and transfer function $T(k)$~\footnote{$n_s$ is one of the six fundamental parameters of the $\Lambda$CDM 
cosmological model~\cite{2020A&A...641A...6P}. $T(k)$ can be expressed in a parametric form and is a function of the matter density~\cite{Eisenstein:1997ik}.}. 
The second approach, utilized in, e.g., Refs.~\cite{Khatri:2024yfr,Sahlu:2023wvl,Kazantzidis:2018rnb} and in our work, 
treats $\sigma_8(z=0)$ as a free parameter for data fitting~\footnote{These two approaches could be regarded as distinct normalization methods 
for LSS ($f\sigma_8$) data. The first method involves normalization at high redshift, whereas the second method employs normalization 
at the present epoch ($z=0$).}. 
Based on current findings, the first method yields smaller errors in constraining $\varepsilon$, 
often resulting in a deviation from zero that exceeds $\sim 3\sigma$~\cite{SolaPeracaula:2017esw}. 
Although the second method appears to lack a comparably strong constraining capability, 
the inclusion of $f\sigma_8$ data can still reduce the errors for constraining $\varepsilon$. 

Additionally, as shown in Ref.~\cite{Pan:2025qwy}, in addition to including DESI BAO (DR2) and CMB data, 
the inclusion of different SNIa datasets produces varying final constraints on $\varepsilon$~\footnote{Note that the LSS ($f\sigma_8$) 
data is not included in their analysis.}. 
For example, some datasets suggest no significant interaction between dark matter and vacuum dark energy, 
while others indicate an interaction scenario at a significance level exceeding $\sim 3\sigma$. 
Regarding the model investigated in this work, although exploring the reasons behind the significant differences 
in constrained results obtained from various SNIa datasets falls outside the scope of this study, it serves as a reminder that, 
at least for the model under consideration, arriving at a definitive and conclusive finding, such as the interaction between dark matter 
and vacuum dark energy, necessitates an extremely cautious approach.


\begin{table*}[htp]

\caption{Constraints on $\varepsilon$ with $1\sigma$ uncertainty level, derived from several previous works that employed diverse datasets for the same 
model currently under consideration. Note that although the abbreviations of some datasets are the same, the data used are different. For example, 
some `CMB' represent complete temperature and polarization data, while others represent distance priors or the shift parameter from CMB data. 
The datasets of BAO and SNIa are also not exactly the same. For specific explanations, one can refer to the main text. 
Furthermore, various data combinations have been employed in some literature to derive distinct limiting results. 
In our citation, we have referenced the most robust limiting result, which is characterized by the smallest error. 
For the findings presented in Ref.~\cite{Pan:2025qwy}, we have referenced the moderate value, which was obtained using the most 
comprehensive SNIa dataset, DESY5. 
}
\label{table:compare}
\begin{center}     
\begin{tabular}{l|c|c}
\toprule

Dataset & $\varepsilon$ & Deviation from $\varepsilon$=0\\
\hline        
DESI BAO+CMB+CC+SNIa+$f\sigma_8$ (this work) &$-0.0073^{+0.0029}_{-0.0033}$ & $2.35\sigma$\\
\hline
BAO+CMB+CC+SNIa+$f\sigma_8$~\cite{SolaPeracaula:2017esw} &$0.0065\pm{0.0018}$ & $3.61\sigma$\\
\hline
Cluster edges+CMB+
BAO+CC+SNIa+GW~\cite{Wang:2020hqq}  &$-0.43^{+0.95}_{-1.30}$ & $0.38\sigma$ \\
\hline
CMB+BAO+CC+SNIa+Lensing \cite{Wang:2018ahw} &$-0.00029^{+0.00028}_{-0.00025}$ & $1.09\sigma$\\
\hline
Galaxy cluster gas mass fraction \cite{SantanaJunior:2024cug} &$-0.191\pm{0.091}$ & $2.10\sigma$  \\
\hline
Angular diameter distance of galaxy cluster \cite{Wang:2017fcr}&$0.0019\pm{0.0089}$ & $0.21\sigma$\\
\hline
HII galaxy \cite{Wang:2016pag} &$0.01^{+0.005}_{-0.007}$ & $1.67\sigma$\\
\hline
BAO+CMB+SNIa+$f\sigma_8$~\cite{Kaeonikhom:2022ahf} &$0.0016\pm{0.0014}$ & $1.14\sigma$\\ 
\hline
DESI BAO+CMB+SNIa~\cite{Li:2024qso} &$-0.0003\pm{0.0011}$ & $0.27\sigma$\\ 
\hline
DESI BAO+CMB+CC+SNIa~\cite{Pan:2025qwy} &$0.0073\pm{0.0035}$ & $2.09\sigma$\\ 

\toprule
\end{tabular}
\end{center}
\end{table*}


\section{conclusions}

We have conducted an investigation into the deviation of dark matter energy density evolution from the standard case, 
denoted as $\rho_{m}\propto (1+z)^{3-\varepsilon}$, where $\varepsilon$ is a constant. This scenario can be interpreted as an 
interaction between dark matter and vacuum dark energy, characterized by an interaction form of $Q=\varepsilon H \rho_{dm}$. 
By utilizing datasets from DESI BAO (DR1), CMB (distance priors from Planck-2018), CC, SNIa (Pantheon), 
and LSS data in the form of the weighted growth rate $f\sigma_8$, we have obtained constraints on $\varepsilon$, 
yielding a value of $-0.0073^{+0.0029}_{-0.0033}$. This finding indicates a deviation of $\sim 2.4\sigma$ from the standard scenario, 
where dark matter and vacuum dark energy do not interact. This conclusion is further supported by the Akaike Information Criterion (AIC). 
Our results confirm those presented in, e.g., Ref.~\cite{SolaPeracaula:2017esw}, 
where the authors found a deviation of around $3.6\sigma$ using different datasets and methodologies. 
The non-zero values of $\varepsilon$ imply the existence of new physics or indicate that dark matter and dark energy possess distinct properties 
compared to those predicted by the $\Lambda$CDM cosmology. 

We also conducted comparisons with several previous studies that utilized different datasets or methodologies. 
In essence, when it comes to the deviation parameter $\varepsilon$, variations in datasets and methodologies often lead to different optimal 
fitting values and associated errors.
Generally, constraints derived from the CMB, SNIa, and previous BAO data do not indicate significant deviation of $\varepsilon$ from zero. 
However, incorporating LSS data, specifically the weighted growth rate measurements $f\sigma_8$, 
can reduce the uncertainties associated with $\varepsilon$ and suggest a deviation from the scenario where dark matter and vacuum dark energy do not interact. 
With the data released by the DESI, $\varepsilon$ exhibits a notable deviation from zero, even when considering combined datasets of 
CMB, SNIa and DESI BAO. 
Nevertheless, it is important to note that, as demonstrated in~\cite{Pan:2025qwy} (where $f\sigma_8$ data was not included), 
different SNIa datasets can have varying impacts on the final constraints 
for $\varepsilon$. Some of these datasets do not show any deviation from zero, even when including data from the second DESI data release. 
This implies that a detailed investigation is needed to understand these differing constraint results for the model under consideration.

\section{Acknowledgements}
We thank Dr. Bin Yue for very useful suggestions. We are grateful to the anonymous referee for his/her very timely and instructive comments 
and suggestions. Additionally, we are deeply appreciative of the editors for their diligent work and for efficiently managing our manuscript 
during the weekend break. This work is supported by the Shandong Provincial Natural Science Foundation (Grant No.ZR2021MA021).

\bibliographystyle{apsrev4-1}
\bibliography{ref}

\begin{thebibliography}{114}%
\makeatletter
\providecommand \@ifxundefined [1]{%
 \@ifx{#1\undefined}
}%
\providecommand \@ifnum [1]{%
 \ifnum #1\expandafter \@firstoftwo
 \else \expandafter \@secondoftwo
 \fi
}%
\providecommand \@ifx [1]{%
 \ifx #1\expandafter \@firstoftwo
 \else \expandafter \@secondoftwo
 \fi
}%
\providecommand \natexlab [1]{#1}%
\providecommand \enquote  [1]{``#1''}%
\providecommand \bibnamefont  [1]{#1}%
\providecommand \bibfnamefont [1]{#1}%
\providecommand \citenamefont [1]{#1}%
\providecommand \href@noop [0]{\@secondoftwo}%
\providecommand \href [0]{\begingroup \@sanitize@url \@href}%
\providecommand \@href[1]{\@@startlink{#1}\@@href}%
\providecommand \@@href[1]{\endgroup#1\@@endlink}%
\providecommand \@sanitize@url [0]{\catcode `\\12\catcode `\$12\catcode
  `\&12\catcode `\#12\catcode `\^12\catcode `\_12\catcode `\%12\relax}%
\providecommand \@@startlink[1]{}%
\providecommand \@@endlink[0]{}%
\providecommand \url  [0]{\begingroup\@sanitize@url \@url }%
\providecommand \@url [1]{\endgroup\@href {#1}{\urlprefix }}%
\providecommand \urlprefix  [0]{URL }%
\providecommand \Eprint [0]{\href }%
\providecommand \doibase [0]{http://dx.doi.org/}%
\providecommand \selectlanguage [0]{\@gobble}%
\providecommand \bibinfo  [0]{\@secondoftwo}%
\providecommand \bibfield  [0]{\@secondoftwo}%
\providecommand \translation [1]{[#1]}%
\providecommand \BibitemOpen [0]{}%
\providecommand \bibitemStop [0]{}%
\providecommand \bibitemNoStop [0]{.\EOS\space}%
\providecommand \EOS [0]{\spacefactor3000\relax}%
\providecommand \BibitemShut  [1]{\csname bibitem#1\endcsname}%
\let\auto@bib@innerbib\@empty
\bibitem [{\citenamefont {Aghanim}\ \emph {et~al.}(2020)\citenamefont {Aghanim}
  \emph {et~al.}}]{2020A&A...641A...6P}%
  \BibitemOpen
  \bibfield  {author} {\bibinfo {author} {\bibfnamefont {N.}~\bibnamefont
  {Aghanim}} \emph {et~al.},\ }\href {\doibase 10.1051/0004-6361/201833910}
  {\bibfield  {journal} {\bibinfo  {journal} {A\&A}\ }\textbf {\bibinfo
  {volume} {641}},\ \bibinfo {eid} {A6} (\bibinfo {year} {2020})},\ \Eprint
  {http://arxiv.org/abs/1807.06209} {arXiv:1807.06209 [astro-ph.CO]}
  \BibitemShut {NoStop}%
\bibitem [{\citenamefont {Copeland}\ \emph {et~al.}(2006)\citenamefont
  {Copeland}, \citenamefont {Sami},\ and\ \citenamefont
  {Tsujikawa}}]{Copeland:2006wr}%
  \BibitemOpen
  \bibfield  {author} {\bibinfo {author} {\bibfnamefont {E.~J.}\ \bibnamefont
  {Copeland}}, \bibinfo {author} {\bibfnamefont {M.}~\bibnamefont {Sami}}, \
  and\ \bibinfo {author} {\bibfnamefont {S.}~\bibnamefont {Tsujikawa}},\ }\href
  {\doibase 10.1142/S021827180600942X} {\bibfield  {journal} {\bibinfo
  {journal} {Int. J. Mod. Phys. D}\ }\textbf {\bibinfo {volume} {15}},\
  \bibinfo {pages} {1753} (\bibinfo {year} {2006})},\ \Eprint
  {http://arxiv.org/abs/hep-th/0603057} {arXiv:hep-th/0603057} \BibitemShut
  {NoStop}%
\bibitem [{\citenamefont {Perivolaropoulos}\ and\ \citenamefont
  {Skara}(2022)}]{Perivolaropoulos:2021jda}%
  \BibitemOpen
  \bibfield  {author} {\bibinfo {author} {\bibfnamefont {L.}~\bibnamefont
  {Perivolaropoulos}}\ and\ \bibinfo {author} {\bibfnamefont {F.}~\bibnamefont
  {Skara}},\ }\href {\doibase 10.1016/j.newar.2022.101659} {\bibfield
  {journal} {\bibinfo  {journal} {New Astron. Rev.}\ }\textbf {\bibinfo
  {volume} {95}},\ \bibinfo {pages} {101659} (\bibinfo {year} {2022})},\
  \Eprint {http://arxiv.org/abs/2105.05208} {arXiv:2105.05208 [astro-ph.CO]}
  \BibitemShut {NoStop}%
\bibitem [{\citenamefont {Hu}\ and\ \citenamefont {Wang}(2023)}]{Hu:2023jqc}%
  \BibitemOpen
  \bibfield  {author} {\bibinfo {author} {\bibfnamefont {J.-P.}\ \bibnamefont
  {Hu}}\ and\ \bibinfo {author} {\bibfnamefont {F.-Y.}\ \bibnamefont {Wang}},\
  }\href {\doibase 10.3390/universe9020094} {\bibfield  {journal} {\bibinfo
  {journal} {Universe}\ }\textbf {\bibinfo {volume} {9}},\ \bibinfo {pages}
  {94} (\bibinfo {year} {2023})},\ \Eprint {http://arxiv.org/abs/2302.05709}
  {arXiv:2302.05709 [astro-ph.CO]} \BibitemShut {NoStop}%
\bibitem [{\citenamefont {Di~Valentino}\ \emph {et~al.}(2021)\citenamefont
  {Di~Valentino}, \citenamefont {Mena}, \citenamefont {Pan}, \citenamefont
  {Visinelli}, \citenamefont {Yang}, \citenamefont {Melchiorri}, \citenamefont
  {Mota}, \citenamefont {Riess},\ and\ \citenamefont
  {Silk}}]{DiValentino:2021izs}%
  \BibitemOpen
  \bibfield  {author} {\bibinfo {author} {\bibfnamefont {E.}~\bibnamefont
  {Di~Valentino}}, \bibinfo {author} {\bibfnamefont {O.}~\bibnamefont {Mena}},
  \bibinfo {author} {\bibfnamefont {S.}~\bibnamefont {Pan}}, \bibinfo {author}
  {\bibfnamefont {L.}~\bibnamefont {Visinelli}}, \bibinfo {author}
  {\bibfnamefont {W.}~\bibnamefont {Yang}}, \bibinfo {author} {\bibfnamefont
  {A.}~\bibnamefont {Melchiorri}}, \bibinfo {author} {\bibfnamefont {D.~F.}\
  \bibnamefont {Mota}}, \bibinfo {author} {\bibfnamefont {A.~G.}\ \bibnamefont
  {Riess}}, \ and\ \bibinfo {author} {\bibfnamefont {J.}~\bibnamefont {Silk}},\
  }\href {\doibase 10.1088/1361-6382/ac086d} {\bibfield  {journal} {\bibinfo
  {journal} {Class. Quant. Grav.}\ }\textbf {\bibinfo {volume} {38}},\ \bibinfo
  {pages} {153001} (\bibinfo {year} {2021})},\ \Eprint
  {http://arxiv.org/abs/2103.01183} {arXiv:2103.01183 [astro-ph.CO]}
  \BibitemShut {NoStop}%
\bibitem [{\citenamefont {Wang}\ \emph {et~al.}(2016)\citenamefont {Wang},
  \citenamefont {Abdalla}, \citenamefont {Atrio-Barandela},\ and\ \citenamefont
  {Pavon}}]{Wang:2016lxa}%
  \BibitemOpen
  \bibfield  {author} {\bibinfo {author} {\bibfnamefont {B.}~\bibnamefont
  {Wang}}, \bibinfo {author} {\bibfnamefont {E.}~\bibnamefont {Abdalla}},
  \bibinfo {author} {\bibfnamefont {F.}~\bibnamefont {Atrio-Barandela}}, \ and\
  \bibinfo {author} {\bibfnamefont {D.}~\bibnamefont {Pavon}},\ }\href
  {\doibase 10.1088/0034-4885/79/9/096901} {\bibfield  {journal} {\bibinfo
  {journal} {Rept. Prog. Phys.}\ }\textbf {\bibinfo {volume} {79}},\ \bibinfo
  {pages} {096901} (\bibinfo {year} {2016})},\ \Eprint
  {http://arxiv.org/abs/1603.08299} {arXiv:1603.08299 [astro-ph.CO]}
  \BibitemShut {NoStop}%
\bibitem [{\citenamefont {Elizalde}\ \emph {et~al.}(2021)\citenamefont
  {Elizalde}, \citenamefont {Gluza},\ and\ \citenamefont
  {Khurshudyan}}]{Elizalde:2021kmo}%
  \BibitemOpen
  \bibfield  {author} {\bibinfo {author} {\bibfnamefont {E.}~\bibnamefont
  {Elizalde}}, \bibinfo {author} {\bibfnamefont {J.}~\bibnamefont {Gluza}}, \
  and\ \bibinfo {author} {\bibfnamefont {M.}~\bibnamefont {Khurshudyan}},\
  }\href@noop {} {\  (\bibinfo {year} {2021})},\ \Eprint
  {http://arxiv.org/abs/2104.01077} {arXiv:2104.01077 [astro-ph.CO]}
  \BibitemShut {NoStop}%
\bibitem [{\citenamefont {Khurshudyan}\ and\ \citenamefont
  {Elizalde}(2024)}]{Khurshudyan:2024gpn}%
  \BibitemOpen
  \bibfield  {author} {\bibinfo {author} {\bibfnamefont {M.}~\bibnamefont
  {Khurshudyan}}\ and\ \bibinfo {author} {\bibfnamefont {E.}~\bibnamefont
  {Elizalde}},\ }\href {\doibase 10.3390/galaxies12040031} {\bibfield
  {journal} {\bibinfo  {journal} {Galaxies}\ }\textbf {\bibinfo {volume}
  {12}},\ \bibinfo {pages} {31} (\bibinfo {year} {2024})},\ \Eprint
  {http://arxiv.org/abs/2402.08630} {arXiv:2402.08630 [gr-qc]} \BibitemShut
  {NoStop}%
\bibitem [{\citenamefont {Khurshudyan}(2023)}]{Khurshudyan:2023cim}%
  \BibitemOpen
  \bibfield  {author} {\bibinfo {author} {\bibfnamefont {M.}~\bibnamefont
  {Khurshudyan}},\ }\href {\doibase 10.1007/s10511-023-09800-3} {\bibfield
  {journal} {\bibinfo  {journal} {Astrophysics}\ }\textbf {\bibinfo {volume}
  {66}},\ \bibinfo {pages} {423} (\bibinfo {year} {2023})},\ \Eprint
  {http://arxiv.org/abs/2308.01233} {arXiv:2308.01233 [gr-qc]} \BibitemShut
  {NoStop}%
\bibitem [{\citenamefont {Kumar}\ \emph {et~al.}(2025)\citenamefont {Kumar},
  \citenamefont {Ajith},\ and\ \citenamefont {Verma}}]{Kumar:2025etf}%
  \BibitemOpen
  \bibfield  {author} {\bibinfo {author} {\bibfnamefont {U.}~\bibnamefont
  {Kumar}}, \bibinfo {author} {\bibfnamefont {A.}~\bibnamefont {Ajith}}, \ and\
  \bibinfo {author} {\bibfnamefont {A.}~\bibnamefont {Verma}},\ }\href@noop {}
  {\  (\bibinfo {year} {2025})},\ \Eprint {http://arxiv.org/abs/2504.14419}
  {arXiv:2504.14419 [astro-ph.CO]} \BibitemShut {NoStop}%
\bibitem [{\citenamefont {Overduin}\ and\ \citenamefont
  {Cooperstock}(1998)}]{Overduin:1998zv}%
  \BibitemOpen
  \bibfield  {author} {\bibinfo {author} {\bibfnamefont {J.~M.}\ \bibnamefont
  {Overduin}}\ and\ \bibinfo {author} {\bibfnamefont {F.~I.}\ \bibnamefont
  {Cooperstock}},\ }\href {\doibase 10.1103/PhysRevD.58.043506} {\bibfield
  {journal} {\bibinfo  {journal} {Phys. Rev. D}\ }\textbf {\bibinfo {volume}
  {58}},\ \bibinfo {pages} {043506} (\bibinfo {year} {1998})},\ \Eprint
  {http://arxiv.org/abs/astro-ph/9805260} {arXiv:astro-ph/9805260} \BibitemShut
  {NoStop}%
\bibitem [{\citenamefont {Carvalho}\ \emph
  {et~al.}(1992{\natexlab{a}})\citenamefont {Carvalho}, \citenamefont {Lima},\
  and\ \citenamefont {Waga}}]{PhysRevD.46.2404}%
  \BibitemOpen
  \bibfield  {author} {\bibinfo {author} {\bibfnamefont {J.~C.}\ \bibnamefont
  {Carvalho}}, \bibinfo {author} {\bibfnamefont {J.~A.~S.}\ \bibnamefont
  {Lima}}, \ and\ \bibinfo {author} {\bibfnamefont {I.}~\bibnamefont {Waga}},\
  }\href {\doibase 10.1103/PhysRevD.46.2404} {\bibfield  {journal} {\bibinfo
  {journal} {Phys. Rev. D}\ }\textbf {\bibinfo {volume} {46}},\ \bibinfo
  {pages} {2404} (\bibinfo {year} {1992}{\natexlab{a}})}\BibitemShut {NoStop}%
\bibitem [{\citenamefont {Arbab}(2003)}]{Arbab:1999hv}%
  \BibitemOpen
  \bibfield  {author} {\bibinfo {author} {\bibfnamefont {A.~I.}\ \bibnamefont
  {Arbab}},\ }\href {\doibase 10.1088/0264-9381/20/1/307} {\bibfield  {journal}
  {\bibinfo  {journal} {Class. Quant. Grav.}\ }\textbf {\bibinfo {volume}
  {20}},\ \bibinfo {pages} {93} (\bibinfo {year} {2003})},\ \Eprint
  {http://arxiv.org/abs/gr-qc/9905066} {arXiv:gr-qc/9905066} \BibitemShut
  {NoStop}%
\bibitem [{\citenamefont {Shapiro}\ and\ \citenamefont
  {Sola}(2002)}]{Shapiro:2000dz}%
  \BibitemOpen
  \bibfield  {author} {\bibinfo {author} {\bibfnamefont {I.~L.}\ \bibnamefont
  {Shapiro}}\ and\ \bibinfo {author} {\bibfnamefont {J.}~\bibnamefont {Sola}},\
  }\href {\doibase 10.1088/1126-6708/2002/02/006} {\bibfield  {journal}
  {\bibinfo  {journal} {JHEP}\ }\textbf {\bibinfo {volume} {02}},\ \bibinfo
  {pages} {006} (\bibinfo {year} {2002})},\ \Eprint
  {http://arxiv.org/abs/hep-th/0012227} {arXiv:hep-th/0012227} \BibitemShut
  {NoStop}%
\bibitem [{\citenamefont {Ray}\ and\ \citenamefont
  {Mukhopadhyay}(2007)}]{Ray:2004nq}%
  \BibitemOpen
  \bibfield  {author} {\bibinfo {author} {\bibfnamefont {S.}~\bibnamefont
  {Ray}}\ and\ \bibinfo {author} {\bibfnamefont {U.}~\bibnamefont
  {Mukhopadhyay}},\ }\href@noop {} {\bibfield  {journal} {\bibinfo  {journal}
  {Grav. Cosmol.}\ }\textbf {\bibinfo {volume} {13}},\ \bibinfo {pages} {142}
  (\bibinfo {year} {2007})},\ \Eprint {http://arxiv.org/abs/astro-ph/0407295}
  {arXiv:astro-ph/0407295} \BibitemShut {NoStop}%
\bibitem [{\citenamefont {Espana-Bonet}\ \emph {et~al.}(2004)\citenamefont
  {Espana-Bonet}, \citenamefont {Ruiz-Lapuente}, \citenamefont {Shapiro},\ and\
  \citenamefont {Sola}}]{Espana-Bonet:2003qjh}%
  \BibitemOpen
  \bibfield  {author} {\bibinfo {author} {\bibfnamefont {C.}~\bibnamefont
  {Espana-Bonet}}, \bibinfo {author} {\bibfnamefont {P.}~\bibnamefont
  {Ruiz-Lapuente}}, \bibinfo {author} {\bibfnamefont {I.~L.}\ \bibnamefont
  {Shapiro}}, \ and\ \bibinfo {author} {\bibfnamefont {J.}~\bibnamefont
  {Sola}},\ }\href {\doibase 10.1088/1475-7516/2004/02/006} {\bibfield
  {journal} {\bibinfo  {journal} {JCAP}\ }\textbf {\bibinfo {volume} {02}},\
  \bibinfo {pages} {006} (\bibinfo {year} {2004})},\ \Eprint
  {http://arxiv.org/abs/hep-ph/0311171} {arXiv:hep-ph/0311171} \BibitemShut
  {NoStop}%
\bibitem [{\citenamefont {Shapiro}\ \emph {et~al.}(2003)\citenamefont
  {Shapiro}, \citenamefont {Sola}, \citenamefont {Espana-Bonet},\ and\
  \citenamefont {Ruiz-Lapuente}}]{Shapiro:2003ui}%
  \BibitemOpen
  \bibfield  {author} {\bibinfo {author} {\bibfnamefont {I.~L.}\ \bibnamefont
  {Shapiro}}, \bibinfo {author} {\bibfnamefont {J.}~\bibnamefont {Sola}},
  \bibinfo {author} {\bibfnamefont {C.}~\bibnamefont {Espana-Bonet}}, \ and\
  \bibinfo {author} {\bibfnamefont {P.}~\bibnamefont {Ruiz-Lapuente}},\ }\href
  {\doibase 10.1016/j.physletb.2003.09.016} {\bibfield  {journal} {\bibinfo
  {journal} {Phys. Lett. B}\ }\textbf {\bibinfo {volume} {574}},\ \bibinfo
  {pages} {149} (\bibinfo {year} {2003})},\ \Eprint
  {http://arxiv.org/abs/astro-ph/0303306} {arXiv:astro-ph/0303306} \BibitemShut
  {NoStop}%
\bibitem [{\citenamefont {G\'omez-Valent}\ \emph {et~al.}(2015)\citenamefont
  {G\'omez-Valent}, \citenamefont {Sol\`a},\ and\ \citenamefont
  {Basilakos}}]{Gomez-Valent:2014rxa}%
  \BibitemOpen
  \bibfield  {author} {\bibinfo {author} {\bibfnamefont {A.}~\bibnamefont
  {G\'omez-Valent}}, \bibinfo {author} {\bibfnamefont {J.}~\bibnamefont
  {Sol\`a}}, \ and\ \bibinfo {author} {\bibfnamefont {S.}~\bibnamefont
  {Basilakos}},\ }\href {\doibase 10.1088/1475-7516/2015/01/004} {\bibfield
  {journal} {\bibinfo  {journal} {JCAP}\ }\textbf {\bibinfo {volume} {01}},\
  \bibinfo {pages} {004} (\bibinfo {year} {2015})},\ \Eprint
  {http://arxiv.org/abs/1409.7048} {arXiv:1409.7048 [astro-ph.CO]} \BibitemShut
  {NoStop}%
\bibitem [{\citenamefont {Sola}\ \emph {et~al.}(2015)\citenamefont {Sola},
  \citenamefont {Gomez-Valent},\ and\ \citenamefont
  {de~Cruz~P\'erez}}]{Sola:2015wwa}%
  \BibitemOpen
  \bibfield  {author} {\bibinfo {author} {\bibfnamefont {J.}~\bibnamefont
  {Sola}}, \bibinfo {author} {\bibfnamefont {A.}~\bibnamefont {Gomez-Valent}},
  \ and\ \bibinfo {author} {\bibfnamefont {J.}~\bibnamefont
  {de~Cruz~P\'erez}},\ }\href {\doibase 10.1088/2041-8205/811/1/L14} {\bibfield
   {journal} {\bibinfo  {journal} {Astrophys. J. Lett.}\ }\textbf {\bibinfo
  {volume} {811}},\ \bibinfo {pages} {L14} (\bibinfo {year} {2015})},\ \Eprint
  {http://arxiv.org/abs/1506.05793} {arXiv:1506.05793 [gr-qc]} \BibitemShut
  {NoStop}%
\bibitem [{\citenamefont {Sol\`a}\ \emph
  {et~al.}(2017{\natexlab{a}})\citenamefont {Sol\`a}, \citenamefont
  {G\'omez-Valent},\ and\ \citenamefont {de~Cruz~P\'erez}}]{Sola:2016jky}%
  \BibitemOpen
  \bibfield  {author} {\bibinfo {author} {\bibfnamefont {J.}~\bibnamefont
  {Sol\`a}}, \bibinfo {author} {\bibfnamefont {A.}~\bibnamefont
  {G\'omez-Valent}}, \ and\ \bibinfo {author} {\bibfnamefont {J.}~\bibnamefont
  {de~Cruz~P\'erez}},\ }\href {\doibase 10.3847/1538-4357/836/1/43} {\bibfield
  {journal} {\bibinfo  {journal} {Astrophys. J.}\ }\textbf {\bibinfo {volume}
  {836}},\ \bibinfo {pages} {43} (\bibinfo {year} {2017}{\natexlab{a}})},\
  \Eprint {http://arxiv.org/abs/1602.02103} {arXiv:1602.02103 [astro-ph.CO]}
  \BibitemShut {NoStop}%
\bibitem [{\citenamefont {Sol\`a~Peracaula}\ \emph {et~al.}(2018)\citenamefont
  {Sol\`a~Peracaula}, \citenamefont {de~Cruz~P\'erez},\ and\ \citenamefont
  {G\'omez-Valent}}]{SolaPeracaula:2016qlq}%
  \BibitemOpen
  \bibfield  {author} {\bibinfo {author} {\bibfnamefont {J.}~\bibnamefont
  {Sol\`a~Peracaula}}, \bibinfo {author} {\bibfnamefont {J.}~\bibnamefont
  {de~Cruz~P\'erez}}, \ and\ \bibinfo {author} {\bibfnamefont {A.}~\bibnamefont
  {G\'omez-Valent}},\ }\href {\doibase 10.1209/0295-5075/121/39001} {\bibfield
  {journal} {\bibinfo  {journal} {EPL}\ }\textbf {\bibinfo {volume} {121}},\
  \bibinfo {pages} {39001} (\bibinfo {year} {2018})},\ \Eprint
  {http://arxiv.org/abs/1606.00450} {arXiv:1606.00450 [gr-qc]} \BibitemShut
  {NoStop}%
\bibitem [{\citenamefont {Sola~Peracaula}\ \emph {et~al.}(2018)\citenamefont
  {Sola~Peracaula}, \citenamefont {de~Cruz~Perez},\ and\ \citenamefont
  {Gomez-Valent}}]{SolaPeracaula:2017esw}%
  \BibitemOpen
  \bibfield  {author} {\bibinfo {author} {\bibfnamefont {J.}~\bibnamefont
  {Sola~Peracaula}}, \bibinfo {author} {\bibfnamefont {J.}~\bibnamefont
  {de~Cruz~Perez}}, \ and\ \bibinfo {author} {\bibfnamefont {A.}~\bibnamefont
  {Gomez-Valent}},\ }\href {\doibase 10.1093/mnras/sty1253} {\bibfield
  {journal} {\bibinfo  {journal} {Mon. Not. Roy. Astron. Soc.}\ }\textbf
  {\bibinfo {volume} {478}},\ \bibinfo {pages} {4357} (\bibinfo {year}
  {2018})},\ \Eprint {http://arxiv.org/abs/1703.08218} {arXiv:1703.08218
  [astro-ph.CO]} \BibitemShut {NoStop}%
\bibitem [{\citenamefont {Sol\`a}\ \emph
  {et~al.}(2017{\natexlab{b}})\citenamefont {Sol\`a}, \citenamefont
  {G\'omez-Valent},\ and\ \citenamefont {de~Cruz~P\'erez}}]{Sola:2017znb}%
  \BibitemOpen
  \bibfield  {author} {\bibinfo {author} {\bibfnamefont {J.}~\bibnamefont
  {Sol\`a}}, \bibinfo {author} {\bibfnamefont {A.}~\bibnamefont
  {G\'omez-Valent}}, \ and\ \bibinfo {author} {\bibfnamefont {J.}~\bibnamefont
  {de~Cruz~P\'erez}},\ }\href {\doibase 10.1016/j.physletb.2017.09.073}
  {\bibfield  {journal} {\bibinfo  {journal} {Phys. Lett. B}\ }\textbf
  {\bibinfo {volume} {774}},\ \bibinfo {pages} {317} (\bibinfo {year}
  {2017}{\natexlab{b}})},\ \Eprint {http://arxiv.org/abs/1705.06723}
  {arXiv:1705.06723 [astro-ph.CO]} \BibitemShut {NoStop}%
\bibitem [{\citenamefont {Yang}\ \emph {et~al.}(2025)\citenamefont {Yang},
  \citenamefont {Wang},\ and\ \citenamefont {Dai}}]{Yang:2025vnm}%
  \BibitemOpen
  \bibfield  {author} {\bibinfo {author} {\bibfnamefont {Y.}~\bibnamefont
  {Yang}}, \bibinfo {author} {\bibfnamefont {Y.}~\bibnamefont {Wang}}, \ and\
  \bibinfo {author} {\bibfnamefont {X.}~\bibnamefont {Dai}},\ }\href {\doibase
  10.1140/epjc/s10052-025-13990-9} {\bibfield  {journal} {\bibinfo  {journal}
  {Eur. Phys. J. C}\ }\textbf {\bibinfo {volume} {85}},\ \bibinfo {pages} {224}
  (\bibinfo {year} {2025})},\ \Eprint {http://arxiv.org/abs/2502.17792}
  {arXiv:2502.17792 [astro-ph.CO]} \BibitemShut {NoStop}%
\bibitem [{\citenamefont {Carvalho}\ \emph
  {et~al.}(1992{\natexlab{b}})\citenamefont {Carvalho}, \citenamefont {Lima},\
  and\ \citenamefont {Waga}}]{PhysRevD462404}%
  \BibitemOpen
  \bibfield  {author} {\bibinfo {author} {\bibfnamefont {J.~C.}\ \bibnamefont
  {Carvalho}}, \bibinfo {author} {\bibfnamefont {J.~A.~S.}\ \bibnamefont
  {Lima}}, \ and\ \bibinfo {author} {\bibfnamefont {I.}~\bibnamefont {Waga}},\
  }\href {\doibase 10.1103/PhysRevD.46.2404} {\bibfield  {journal} {\bibinfo
  {journal} {Phys. Rev. D}\ }\textbf {\bibinfo {volume} {46}},\ \bibinfo
  {pages} {2404} (\bibinfo {year} {1992}{\natexlab{b}})}\BibitemShut {NoStop}%
\bibitem [{\citenamefont {Vishwakarma}(2000)}]{RGVishwakarma_2000}%
  \BibitemOpen
  \bibfield  {author} {\bibinfo {author} {\bibfnamefont {R.~G.}\ \bibnamefont
  {Vishwakarma}},\ }\href {\doibase 10.1088/0264-9381/17/18/317} {\bibfield
  {journal} {\bibinfo  {journal} {Classical and Quantum Gravity}\ }\textbf
  {\bibinfo {volume} {17}},\ \bibinfo {pages} {3833} (\bibinfo {year}
  {2000})}\BibitemShut {NoStop}%
\bibitem [{\citenamefont {Kaeonikhom}\ \emph {et~al.}(2023)\citenamefont
  {Kaeonikhom}, \citenamefont {Assadullahi}, \citenamefont {Schewtschenko},\
  and\ \citenamefont {Wands}}]{Kaeonikhom:2022ahf}%
  \BibitemOpen
  \bibfield  {author} {\bibinfo {author} {\bibfnamefont {C.}~\bibnamefont
  {Kaeonikhom}}, \bibinfo {author} {\bibfnamefont {H.}~\bibnamefont
  {Assadullahi}}, \bibinfo {author} {\bibfnamefont {J.}~\bibnamefont
  {Schewtschenko}}, \ and\ \bibinfo {author} {\bibfnamefont {D.}~\bibnamefont
  {Wands}},\ }\href {\doibase 10.1088/1475-7516/2023/01/042} {\bibfield
  {journal} {\bibinfo  {journal} {JCAP}\ }\textbf {\bibinfo {volume} {01}},\
  \bibinfo {pages} {042} (\bibinfo {year} {2023})},\ \Eprint
  {http://arxiv.org/abs/2210.05363} {arXiv:2210.05363 [astro-ph.CO]}
  \BibitemShut {NoStop}%
\bibitem [{\citenamefont {{Doroshkevich}}\ and\ \citenamefont
  {{Khlopov}}(1984)}]{1984MNRAS.211..277D}%
  \BibitemOpen
  \bibfield  {author} {\bibinfo {author} {\bibfnamefont {A.~G.}\ \bibnamefont
  {{Doroshkevich}}}\ and\ \bibinfo {author} {\bibfnamefont {M.~I.}\
  \bibnamefont {{Khlopov}}},\ }\href {\doibase 10.1093/mnras/211.2.277}
  {\bibfield  {journal} {\bibinfo  {journal} {MNRAS}\ }\textbf {\bibinfo
  {volume} {211}},\ \bibinfo {pages} {277} (\bibinfo {year}
  {1984})}\BibitemShut {NoStop}%
\bibitem [{\citenamefont {Doroshkevich}\ and\ \citenamefont
  {Khlopov}(1985)}]{Doroshkevich}%
  \BibitemOpen
  \bibfield  {author} {\bibinfo {author} {\bibfnamefont {A.}~\bibnamefont
  {Doroshkevich}}\ and\ \bibinfo {author} {\bibfnamefont {M.}~\bibnamefont
  {Khlopov}},\ }\href@noop {} {\bibfield  {journal} {\bibinfo  {journal}
  {Soviet Astronomy Letters}\ }\textbf {\bibinfo {volume} {11}},\ \bibinfo
  {pages} {236} (\bibinfo {year} {1985})}\BibitemShut {NoStop}%
\bibitem [{\citenamefont {Doroshkevich}\ and\ \citenamefont
  {Khlopov}(1988)}]{cmun}%
  \BibitemOpen
  \bibfield  {author} {\bibinfo {author} {\bibfnamefont {A.}~\bibnamefont
  {Doroshkevich}}\ and\ \bibinfo {author} {\bibfnamefont {M.}~\bibnamefont
  {Khlopov}},\ }\href@noop {} {\bibfield  {journal} {\bibinfo  {journal}
  {Soviet Astronomy}\ }\textbf {\bibinfo {volume} {32}},\ \bibinfo {pages}
  {127} (\bibinfo {year} {1988})}\BibitemShut {NoStop}%
\bibitem [{\citenamefont {Doroshkevich}\ \emph {et~al.}(1989)\citenamefont
  {Doroshkevich}, \citenamefont {Klypin},\ and\ \citenamefont
  {Khlopov}}]{10.1093/mnras/239.3.923}%
  \BibitemOpen
  \bibfield  {author} {\bibinfo {author} {\bibfnamefont {A.~G.}\ \bibnamefont
  {Doroshkevich}}, \bibinfo {author} {\bibfnamefont {A.~A.}\ \bibnamefont
  {Klypin}}, \ and\ \bibinfo {author} {\bibfnamefont {M.~U.}\ \bibnamefont
  {Khlopov}},\ }\href {\doibase 10.1093/mnras/239.3.923} {\bibfield  {journal}
  {\bibinfo  {journal} {Monthly Notices of the Royal Astronomical Society}\
  }\textbf {\bibinfo {volume} {239}},\ \bibinfo {pages} {923} (\bibinfo {year}
  {1989})}\BibitemShut {NoStop}%
\bibitem [{\citenamefont {Doroshkevich}\ and\ \citenamefont
  {Khlopov}(1984)}]{Doroshkevich:1984gw}%
  \BibitemOpen
  \bibfield  {author} {\bibinfo {author} {\bibfnamefont {A.~G.}\ \bibnamefont
  {Doroshkevich}}\ and\ \bibinfo {author} {\bibfnamefont {M.~Y.}\ \bibnamefont
  {Khlopov}},\ }\href@noop {} {\bibfield  {journal} {\bibinfo  {journal} {Yad.
  Fiz.}\ }\textbf {\bibinfo {volume} {39}},\ \bibinfo {pages} {869} (\bibinfo
  {year} {1984})}\BibitemShut {NoStop}%
\bibitem [{\citenamefont {Wu}(2025)}]{Wu:2025wyk}%
  \BibitemOpen
  \bibfield  {author} {\bibinfo {author} {\bibfnamefont {P.-J.}\ \bibnamefont
  {Wu}},\ }\href@noop {} {\  (\bibinfo {year} {2025})},\ \Eprint
  {http://arxiv.org/abs/2504.09054} {arXiv:2504.09054 [astro-ph.CO]}
  \BibitemShut {NoStop}%
\bibitem [{\citenamefont {Wu}\ and\ \citenamefont {Zhang}(2024)}]{Wu:2024faw}%
  \BibitemOpen
  \bibfield  {author} {\bibinfo {author} {\bibfnamefont {P.-J.}\ \bibnamefont
  {Wu}}\ and\ \bibinfo {author} {\bibfnamefont {X.}~\bibnamefont {Zhang}},\
  }\href@noop {} {\  (\bibinfo {year} {2024})},\ \Eprint
  {http://arxiv.org/abs/2411.06356} {arXiv:2411.06356 [astro-ph.CO]}
  \BibitemShut {NoStop}%
\bibitem [{\citenamefont {Wang}\ and\ \citenamefont
  {Meng}(2005)}]{Wang:2004cp}%
  \BibitemOpen
  \bibfield  {author} {\bibinfo {author} {\bibfnamefont {P.}~\bibnamefont
  {Wang}}\ and\ \bibinfo {author} {\bibfnamefont {X.-H.}\ \bibnamefont
  {Meng}},\ }\href {\doibase 10.1088/0264-9381/22/2/003} {\bibfield  {journal}
  {\bibinfo  {journal} {Class. Quant. Grav.}\ }\textbf {\bibinfo {volume}
  {22}},\ \bibinfo {pages} {283} (\bibinfo {year} {2005})},\ \Eprint
  {http://arxiv.org/abs/astro-ph/0408495} {arXiv:astro-ph/0408495} \BibitemShut
  {NoStop}%
\bibitem [{\citenamefont {Santana~J\'unior}\ \emph {et~al.}(2024)\citenamefont
  {Santana~J\'unior}, \citenamefont {Costa}, \citenamefont {Holanda},\ and\
  \citenamefont {Silva}}]{SantanaJunior:2024cug}%
  \BibitemOpen
  \bibfield  {author} {\bibinfo {author} {\bibfnamefont {Z.~C.}\ \bibnamefont
  {Santana~J\'unior}}, \bibinfo {author} {\bibfnamefont {M.~O.}\ \bibnamefont
  {Costa}}, \bibinfo {author} {\bibfnamefont {R.~F.~L.}\ \bibnamefont
  {Holanda}}, \ and\ \bibinfo {author} {\bibfnamefont {R.}~\bibnamefont
  {Silva}},\ }\href {\doibase 10.1103/PhysRevD.109.123542} {\bibfield
  {journal} {\bibinfo  {journal} {Phys. Rev. D}\ }\textbf {\bibinfo {volume}
  {109}},\ \bibinfo {pages} {123542} (\bibinfo {year} {2024})},\ \Eprint
  {http://arxiv.org/abs/2405.15726} {arXiv:2405.15726 [astro-ph.CO]}
  \BibitemShut {NoStop}%
\bibitem [{\citenamefont {Koussour}\ \emph {et~al.}(2024)\citenamefont
  {Koussour}, \citenamefont {Myrzakulov},\ and\ \citenamefont
  {Rayimbaev}}]{Koussour:2024nhw}%
  \BibitemOpen
  \bibfield  {author} {\bibinfo {author} {\bibfnamefont {M.}~\bibnamefont
  {Koussour}}, \bibinfo {author} {\bibfnamefont {N.}~\bibnamefont
  {Myrzakulov}}, \ and\ \bibinfo {author} {\bibfnamefont {J.}~\bibnamefont
  {Rayimbaev}},\ }\href {\doibase 10.1016/j.asr.2024.04.045} {\bibfield
  {journal} {\bibinfo  {journal} {Adv. Space Res.}\ }\textbf {\bibinfo {volume}
  {74}},\ \bibinfo {pages} {1343} (\bibinfo {year} {2024})},\ \Eprint
  {http://arxiv.org/abs/2404.15982} {arXiv:2404.15982 [astro-ph.CO]}
  \BibitemShut {NoStop}%
\bibitem [{\citenamefont {\"Ozta\c{s}}(2018)}]{Oztas:2018zvi}%
  \BibitemOpen
  \bibfield  {author} {\bibinfo {author} {\bibfnamefont {A.~M.}\ \bibnamefont
  {\"Ozta\c{s}}},\ }\href {\doibase 10.1093/mnras/sty2375} {\bibfield
  {journal} {\bibinfo  {journal} {Mon. Not. Roy. Astron. Soc.}\ }\textbf
  {\bibinfo {volume} {481}},\ \bibinfo {pages} {2228} (\bibinfo {year}
  {2018})}\BibitemShut {NoStop}%
\bibitem [{\citenamefont {Vishwakarma}(2001)}]{Vishwakarma:2000gp}%
  \BibitemOpen
  \bibfield  {author} {\bibinfo {author} {\bibfnamefont {R.~G.}\ \bibnamefont
  {Vishwakarma}},\ }\href {\doibase 10.1088/0264-9381/18/7/301} {\bibfield
  {journal} {\bibinfo  {journal} {Class. Quant. Grav.}\ }\textbf {\bibinfo
  {volume} {18}},\ \bibinfo {pages} {1159} (\bibinfo {year} {2001})},\ \Eprint
  {http://arxiv.org/abs/astro-ph/0012492} {arXiv:astro-ph/0012492} \BibitemShut
  {NoStop}%
\bibitem [{\citenamefont {Azri}\ and\ \citenamefont
  {Bounames}(2017)}]{Azri:2014apa}%
  \BibitemOpen
  \bibfield  {author} {\bibinfo {author} {\bibfnamefont {H.}~\bibnamefont
  {Azri}}\ and\ \bibinfo {author} {\bibfnamefont {A.}~\bibnamefont
  {Bounames}},\ }\href {\doibase 10.1142/S0218271817500602} {\bibfield
  {journal} {\bibinfo  {journal} {Int. J. Mod. Phys. D}\ }\textbf {\bibinfo
  {volume} {26}},\ \bibinfo {pages} {1750060} (\bibinfo {year} {2017})},\
  \Eprint {http://arxiv.org/abs/1412.7567} {arXiv:1412.7567 [gr-qc]}
  \BibitemShut {NoStop}%
\bibitem [{\citenamefont {Azri}\ and\ \citenamefont
  {Bounames}(2012)}]{Azri:2012fpi}%
  \BibitemOpen
  \bibfield  {author} {\bibinfo {author} {\bibfnamefont {H.}~\bibnamefont
  {Azri}}\ and\ \bibinfo {author} {\bibfnamefont {A.}~\bibnamefont
  {Bounames}},\ }\href {\doibase 10.1007/s10714-012-1413-9} {\bibfield
  {journal} {\bibinfo  {journal} {Gen. Rel. Grav.}\ }\textbf {\bibinfo {volume}
  {44}},\ \bibinfo {pages} {2547} (\bibinfo {year} {2012})},\ \Eprint
  {http://arxiv.org/abs/1007.1948} {arXiv:1007.1948 [gr-qc]} \BibitemShut
  {NoStop}%
\bibitem [{\citenamefont {Szydowski}(2015)}]{Szydlowski:2015bwa}%
  \BibitemOpen
  \bibfield  {author} {\bibinfo {author} {\bibfnamefont {M.}~\bibnamefont
  {Szydowski}},\ }\href {\doibase 10.1103/PhysRevD.91.123538} {\bibfield
  {journal} {\bibinfo  {journal} {Phys. Rev. D}\ }\textbf {\bibinfo {volume}
  {91}},\ \bibinfo {pages} {123538} (\bibinfo {year} {2015})},\ \Eprint
  {http://arxiv.org/abs/1502.04737} {arXiv:1502.04737 [astro-ph.CO]}
  \BibitemShut {NoStop}%
\bibitem [{\citenamefont {Bruni}\ \emph {et~al.}(2022)\citenamefont {Bruni},
  \citenamefont {Maier},\ and\ \citenamefont {Wands}}]{Bruni:2021msx}%
  \BibitemOpen
  \bibfield  {author} {\bibinfo {author} {\bibfnamefont {M.}~\bibnamefont
  {Bruni}}, \bibinfo {author} {\bibfnamefont {R.}~\bibnamefont {Maier}}, \ and\
  \bibinfo {author} {\bibfnamefont {D.}~\bibnamefont {Wands}},\ }\href
  {\doibase 10.1103/PhysRevD.105.063532} {\bibfield  {journal} {\bibinfo
  {journal} {Phys. Rev. D}\ }\textbf {\bibinfo {volume} {105}},\ \bibinfo
  {pages} {063532} (\bibinfo {year} {2022})},\ \Eprint
  {http://arxiv.org/abs/2111.01765} {arXiv:2111.01765 [gr-qc]} \BibitemShut
  {NoStop}%
\bibitem [{\citenamefont {Sola~Peracaula}\ \emph {et~al.}(2023)\citenamefont
  {Sola~Peracaula}, \citenamefont {Gomez-Valent}, \citenamefont
  {de~Cruz~Perez},\ and\ \citenamefont
  {Moreno-Pulido}}]{SolaPeracaula:2023swx}%
  \BibitemOpen
  \bibfield  {author} {\bibinfo {author} {\bibfnamefont {J.}~\bibnamefont
  {Sola~Peracaula}}, \bibinfo {author} {\bibfnamefont {A.}~\bibnamefont
  {Gomez-Valent}}, \bibinfo {author} {\bibfnamefont {J.}~\bibnamefont
  {de~Cruz~Perez}}, \ and\ \bibinfo {author} {\bibfnamefont {C.}~\bibnamefont
  {Moreno-Pulido}},\ }\href {\doibase 10.3390/universe9060262} {\bibfield
  {journal} {\bibinfo  {journal} {Universe}\ }\textbf {\bibinfo {volume} {9}},\
  \bibinfo {pages} {262} (\bibinfo {year} {2023})},\ \Eprint
  {http://arxiv.org/abs/2304.11157} {arXiv:2304.11157 [astro-ph.CO]}
  \BibitemShut {NoStop}%
\bibitem [{\citenamefont {Bora}\ \emph {et~al.}(2021)\citenamefont {Bora},
  \citenamefont {Holanda},\ and\ \citenamefont {Desai}}]{Bora:2021uxq}%
  \BibitemOpen
  \bibfield  {author} {\bibinfo {author} {\bibfnamefont {K.}~\bibnamefont
  {Bora}}, \bibinfo {author} {\bibfnamefont {R.~F.~L.}\ \bibnamefont
  {Holanda}}, \ and\ \bibinfo {author} {\bibfnamefont {S.}~\bibnamefont
  {Desai}},\ }\href {\doibase 10.1140/epjc/s10052-021-09421-0} {\bibfield
  {journal} {\bibinfo  {journal} {Eur. Phys. J. C}\ }\textbf {\bibinfo {volume}
  {81}},\ \bibinfo {pages} {596} (\bibinfo {year} {2021})},\ \Eprint
  {http://arxiv.org/abs/2105.02168} {arXiv:2105.02168 [astro-ph.CO]}
  \BibitemShut {NoStop}%
\bibitem [{\citenamefont {Bora}\ \emph {et~al.}(2022)\citenamefont {Bora},
  \citenamefont {Holanda}, \citenamefont {Desai},\ and\ \citenamefont
  {Pereira}}]{Bora:2021iww}%
  \BibitemOpen
  \bibfield  {author} {\bibinfo {author} {\bibfnamefont {K.}~\bibnamefont
  {Bora}}, \bibinfo {author} {\bibfnamefont {R.~F.~L.}\ \bibnamefont
  {Holanda}}, \bibinfo {author} {\bibfnamefont {S.}~\bibnamefont {Desai}}, \
  and\ \bibinfo {author} {\bibfnamefont {S.~H.}\ \bibnamefont {Pereira}},\
  }\href {\doibase 10.1140/epjc/s10052-022-09987-3} {\bibfield  {journal}
  {\bibinfo  {journal} {Eur. Phys. J. C}\ }\textbf {\bibinfo {volume} {82}},\
  \bibinfo {pages} {17} (\bibinfo {year} {2022})},\ \Eprint
  {http://arxiv.org/abs/2106.15805} {arXiv:2106.15805 [astro-ph.CO]}
  \BibitemShut {NoStop}%
\bibitem [{\citenamefont {Holanda}\ \emph {et~al.}(2019)\citenamefont
  {Holanda}, \citenamefont {Gon\c{c}alves}, \citenamefont {Gonzalez},\ and\
  \citenamefont {Alcaniz}}]{Holanda:2019sod}%
  \BibitemOpen
  \bibfield  {author} {\bibinfo {author} {\bibfnamefont {R.~F.~L.}\
  \bibnamefont {Holanda}}, \bibinfo {author} {\bibfnamefont {R.~S.}\
  \bibnamefont {Gon\c{c}alves}}, \bibinfo {author} {\bibfnamefont {J.~E.}\
  \bibnamefont {Gonzalez}}, \ and\ \bibinfo {author} {\bibfnamefont {J.~S.}\
  \bibnamefont {Alcaniz}},\ }\href {\doibase 10.1088/1475-7516/2019/11/032}
  {\bibfield  {journal} {\bibinfo  {journal} {JCAP}\ }\textbf {\bibinfo
  {volume} {11}},\ \bibinfo {pages} {032} (\bibinfo {year} {2019})},\ \Eprint
  {http://arxiv.org/abs/1905.09689} {arXiv:1905.09689 [astro-ph.CO]}
  \BibitemShut {NoStop}%
\bibitem [{\citenamefont {Brito}\ \emph {et~al.}(2024)\citenamefont {Brito},
  \citenamefont {Jesus}, \citenamefont {Escobal},\ and\ \citenamefont
  {Pereira}}]{Brito:2024bhh}%
  \BibitemOpen
  \bibfield  {author} {\bibinfo {author} {\bibfnamefont {L.~S.}\ \bibnamefont
  {Brito}}, \bibinfo {author} {\bibfnamefont {J.~F.}\ \bibnamefont {Jesus}},
  \bibinfo {author} {\bibfnamefont {A.~A.}\ \bibnamefont {Escobal}}, \ and\
  \bibinfo {author} {\bibfnamefont {S.~H.}\ \bibnamefont {Pereira}},\
  }\href@noop {} {\  (\bibinfo {year} {2024})},\ \Eprint
  {http://arxiv.org/abs/2412.06756} {arXiv:2412.06756 [astro-ph.CO]}
  \BibitemShut {NoStop}%
\bibitem [{\citenamefont {Feng}\ \emph {et~al.}(2020)\citenamefont {Feng},
  \citenamefont {He}, \citenamefont {Li}, \citenamefont {Zhang},\ and\
  \citenamefont {Zhang}}]{Feng:2019jqa}%
  \BibitemOpen
  \bibfield  {author} {\bibinfo {author} {\bibfnamefont {L.}~\bibnamefont
  {Feng}}, \bibinfo {author} {\bibfnamefont {D.-Z.}\ \bibnamefont {He}},
  \bibinfo {author} {\bibfnamefont {H.-L.}\ \bibnamefont {Li}}, \bibinfo
  {author} {\bibfnamefont {J.-F.}\ \bibnamefont {Zhang}}, \ and\ \bibinfo
  {author} {\bibfnamefont {X.}~\bibnamefont {Zhang}},\ }\href {\doibase
  10.1007/s11433-019-1511-8} {\bibfield  {journal} {\bibinfo  {journal} {Sci.
  China Phys. Mech. Astron.}\ }\textbf {\bibinfo {volume} {63}},\ \bibinfo
  {pages} {290404} (\bibinfo {year} {2020})},\ \Eprint
  {http://arxiv.org/abs/1910.03872} {arXiv:1910.03872 [astro-ph.CO]}
  \BibitemShut {NoStop}%
\bibitem [{\citenamefont {Barboza}\ \emph {et~al.}(2024)\citenamefont
  {Barboza}, \citenamefont {Oliveira},\ and\ \citenamefont
  {Soares}}]{Barboza:2024gcd}%
  \BibitemOpen
  \bibfield  {author} {\bibinfo {author} {\bibfnamefont {E.~M.}\ \bibnamefont
  {Barboza}}, \bibinfo {author} {\bibfnamefont {L.~J.~S.}\ \bibnamefont
  {Oliveira}}, \ and\ \bibinfo {author} {\bibfnamefont {B.~B.}\ \bibnamefont
  {Soares}},\ }\href {\doibase 10.1142/S021827182450010X} {\bibfield  {journal}
  {\bibinfo  {journal} {Int. J. Mod. Phys. D}\ }\textbf {\bibinfo {volume}
  {33}},\ \bibinfo {pages} {2450010} (\bibinfo {year} {2024})}\BibitemShut
  {NoStop}%
\bibitem [{\citenamefont {Basilakos}\ and\ \citenamefont
  {Sol\`a}(2014)}]{PhysRevD.90.023008}%
  \BibitemOpen
  \bibfield  {author} {\bibinfo {author} {\bibfnamefont {S.}~\bibnamefont
  {Basilakos}}\ and\ \bibinfo {author} {\bibfnamefont {J.}~\bibnamefont
  {Sol\`a}},\ }\href {\doibase 10.1103/PhysRevD.90.023008} {\bibfield
  {journal} {\bibinfo  {journal} {Phys. Rev. D}\ }\textbf {\bibinfo {volume}
  {90}},\ \bibinfo {pages} {023008} (\bibinfo {year} {2014})}\BibitemShut
  {NoStop}%
\bibitem [{\citenamefont {G\'omez-Valent}\ and\ \citenamefont
  {Sol\`a~Peracaula}(2018)}]{Gomez-Valent:2018nib}%
  \BibitemOpen
  \bibfield  {author} {\bibinfo {author} {\bibfnamefont {A.}~\bibnamefont
  {G\'omez-Valent}}\ and\ \bibinfo {author} {\bibfnamefont {J.}~\bibnamefont
  {Sol\`a~Peracaula}},\ }\href {\doibase 10.1093/mnras/sty1028} {\bibfield
  {journal} {\bibinfo  {journal} {Mon. Not. Roy. Astron. Soc.}\ }\textbf
  {\bibinfo {volume} {478}},\ \bibinfo {pages} {126} (\bibinfo {year}
  {2018})},\ \Eprint {http://arxiv.org/abs/1801.08501} {arXiv:1801.08501
  [astro-ph.CO]} \BibitemShut {NoStop}%
\bibitem [{\citenamefont {Adame}\ \emph {et~al.}(2024)\citenamefont {Adame}
  \emph {et~al.}}]{DESI:2024mwx}%
  \BibitemOpen
  \bibfield  {author} {\bibinfo {author} {\bibfnamefont {A.~G.}\ \bibnamefont
  {Adame}} \emph {et~al.} (\bibinfo {collaboration} {DESI}),\ }\href@noop {} {\
   (\bibinfo {year} {2024})},\ \Eprint {http://arxiv.org/abs/2404.03002}
  {arXiv:2404.03002 [astro-ph.CO]} \BibitemShut {NoStop}%
\bibitem [{\citenamefont {Li}\ and\ \citenamefont {Wang}(2024)}]{Li:2024hrv}%
  \BibitemOpen
  \bibfield  {author} {\bibinfo {author} {\bibfnamefont {J.-X.}\ \bibnamefont
  {Li}}\ and\ \bibinfo {author} {\bibfnamefont {S.}~\bibnamefont {Wang}},\
  }\href@noop {} {\  (\bibinfo {year} {2024})},\ \Eprint
  {http://arxiv.org/abs/2412.09064} {arXiv:2412.09064 [astro-ph.CO]}
  \BibitemShut {NoStop}%
\bibitem [{\citenamefont {Chen}\ \emph {et~al.}(2019)\citenamefont {Chen},
  \citenamefont {Huang},\ and\ \citenamefont {Wang}}]{Chen:2018dbv}%
  \BibitemOpen
  \bibfield  {author} {\bibinfo {author} {\bibfnamefont {L.}~\bibnamefont
  {Chen}}, \bibinfo {author} {\bibfnamefont {Q.-G.}\ \bibnamefont {Huang}}, \
  and\ \bibinfo {author} {\bibfnamefont {K.}~\bibnamefont {Wang}},\ }\href
  {\doibase 10.1088/1475-7516/2019/02/028} {\bibfield  {journal} {\bibinfo
  {journal} {JCAP}\ }\textbf {\bibinfo {volume} {02}},\ \bibinfo {pages} {028}
  (\bibinfo {year} {2019})},\ \Eprint {http://arxiv.org/abs/1808.05724}
  {arXiv:1808.05724 [astro-ph.CO]} \BibitemShut {NoStop}%
\bibitem [{\citenamefont {Liu}\ \emph {et~al.}(2019)\citenamefont {Liu},
  \citenamefont {Guo}, \citenamefont {Zhang},\ and\ \citenamefont
  {Zhang}}]{Liu:2018kjv}%
  \BibitemOpen
  \bibfield  {author} {\bibinfo {author} {\bibfnamefont {Y.}~\bibnamefont
  {Liu}}, \bibinfo {author} {\bibfnamefont {R.-Y.}\ \bibnamefont {Guo}},
  \bibinfo {author} {\bibfnamefont {J.-F.}\ \bibnamefont {Zhang}}, \ and\
  \bibinfo {author} {\bibfnamefont {X.}~\bibnamefont {Zhang}},\ }\href
  {\doibase 10.1088/1475-7516/2019/05/016} {\bibfield  {journal} {\bibinfo
  {journal} {JCAP}\ }\textbf {\bibinfo {volume} {05}},\ \bibinfo {pages} {016}
  (\bibinfo {year} {2019})},\ \Eprint {http://arxiv.org/abs/1811.12131}
  {arXiv:1811.12131 [astro-ph.CO]} \BibitemShut {NoStop}%
\bibitem [{\citenamefont {Xu}\ and\ \citenamefont {Zhang}(2016)}]{Xu:2016grp}%
  \BibitemOpen
  \bibfield  {author} {\bibinfo {author} {\bibfnamefont {Y.-Y.}\ \bibnamefont
  {Xu}}\ and\ \bibinfo {author} {\bibfnamefont {X.}~\bibnamefont {Zhang}},\
  }\href {\doibase 10.1140/epjc/s10052-016-4446-5} {\bibfield  {journal}
  {\bibinfo  {journal} {Eur. Phys. J. C}\ }\textbf {\bibinfo {volume} {76}},\
  \bibinfo {pages} {588} (\bibinfo {year} {2016})},\ \Eprint
  {http://arxiv.org/abs/1607.06262} {arXiv:1607.06262 [astro-ph.CO]}
  \BibitemShut {NoStop}%
\bibitem [{\citenamefont {Feng}\ and\ \citenamefont {Lu}(2011)}]{Feng_2011}%
  \BibitemOpen
  \bibfield  {author} {\bibinfo {author} {\bibfnamefont {L.}~\bibnamefont
  {Feng}}\ and\ \bibinfo {author} {\bibfnamefont {T.}~\bibnamefont {Lu}},\
  }\href {\doibase 10.1088/1475-7516/2011/11/034} {\bibfield  {journal}
  {\bibinfo  {journal} {Journal of Cosmology and Astroparticle Physics}\
  }\textbf {\bibinfo {volume} {2011}},\ \bibinfo {pages} {034–034} (\bibinfo
  {year} {2011})}\BibitemShut {NoStop}%
\bibitem [{\citenamefont {Hu}\ and\ \citenamefont
  {Sugiyama}(1996)}]{Hu:1995en}%
  \BibitemOpen
  \bibfield  {author} {\bibinfo {author} {\bibfnamefont {W.}~\bibnamefont
  {Hu}}\ and\ \bibinfo {author} {\bibfnamefont {N.}~\bibnamefont {Sugiyama}},\
  }\href {\doibase 10.1086/177989} {\bibfield  {journal} {\bibinfo  {journal}
  {Astrophys. J.}\ }\textbf {\bibinfo {volume} {471}},\ \bibinfo {pages} {542}
  (\bibinfo {year} {1996})},\ \Eprint {http://arxiv.org/abs/astro-ph/9510117}
  {arXiv:astro-ph/9510117} \BibitemShut {NoStop}%
\bibitem [{\citenamefont {Zhai}\ and\ \citenamefont
  {Wang}(2019)}]{Zhai:2018vmm}%
  \BibitemOpen
  \bibfield  {author} {\bibinfo {author} {\bibfnamefont {Z.}~\bibnamefont
  {Zhai}}\ and\ \bibinfo {author} {\bibfnamefont {Y.}~\bibnamefont {Wang}},\
  }\href {\doibase 10.1088/1475-7516/2019/07/005} {\bibfield  {journal}
  {\bibinfo  {journal} {JCAP}\ }\textbf {\bibinfo {volume} {07}},\ \bibinfo
  {pages} {005} (\bibinfo {year} {2019})},\ \Eprint
  {http://arxiv.org/abs/1811.07425} {arXiv:1811.07425 [astro-ph.CO]}
  \BibitemShut {NoStop}%
\bibitem [{\citenamefont {Jimenez}\ and\ \citenamefont
  {Loeb}(2002)}]{Jimenez:2001gg}%
  \BibitemOpen
  \bibfield  {author} {\bibinfo {author} {\bibfnamefont {R.}~\bibnamefont
  {Jimenez}}\ and\ \bibinfo {author} {\bibfnamefont {A.}~\bibnamefont {Loeb}},\
  }\href {\doibase 10.1086/340549} {\bibfield  {journal} {\bibinfo  {journal}
  {Astrophys. J.}\ }\textbf {\bibinfo {volume} {573}},\ \bibinfo {pages} {37}
  (\bibinfo {year} {2002})},\ \Eprint {http://arxiv.org/abs/astro-ph/0106145}
  {arXiv:astro-ph/0106145} \BibitemShut {NoStop}%
\bibitem [{\citenamefont {Moresco}\ \emph {et~al.}(2022)\citenamefont {Moresco}
  \emph {et~al.}}]{Moresco:2022phi}%
  \BibitemOpen
  \bibfield  {author} {\bibinfo {author} {\bibfnamefont {M.}~\bibnamefont
  {Moresco}} \emph {et~al.},\ }\href {\doibase 10.1007/s41114-022-00040-z}
  {\bibfield  {journal} {\bibinfo  {journal} {Living Rev. Rel.}\ }\textbf
  {\bibinfo {volume} {25}},\ \bibinfo {pages} {6} (\bibinfo {year} {2022})},\
  \Eprint {http://arxiv.org/abs/2201.07241} {arXiv:2201.07241 [astro-ph.CO]}
  \BibitemShut {NoStop}%
\bibitem [{\citenamefont {Jimenez}\ \emph {et~al.}(2023)\citenamefont
  {Jimenez}, \citenamefont {Moresco}, \citenamefont {Verde},\ and\
  \citenamefont {Wandelt}}]{Jimenez:2023flo}%
  \BibitemOpen
  \bibfield  {author} {\bibinfo {author} {\bibfnamefont {R.}~\bibnamefont
  {Jimenez}}, \bibinfo {author} {\bibfnamefont {M.}~\bibnamefont {Moresco}},
  \bibinfo {author} {\bibfnamefont {L.}~\bibnamefont {Verde}}, \ and\ \bibinfo
  {author} {\bibfnamefont {B.~D.}\ \bibnamefont {Wandelt}},\ }\href {\doibase
  10.1088/1475-7516/2023/11/047} {\bibfield  {journal} {\bibinfo  {journal}
  {JCAP}\ }\textbf {\bibinfo {volume} {11}},\ \bibinfo {pages} {047} (\bibinfo
  {year} {2023})},\ \Eprint {http://arxiv.org/abs/2306.11425} {arXiv:2306.11425
  [astro-ph.CO]} \BibitemShut {NoStop}%
\bibitem [{\citenamefont {Li}\ \emph {et~al.}(2021)\citenamefont {Li},
  \citenamefont {Du}, \citenamefont {Zhou}, \citenamefont {Zhang},\ and\
  \citenamefont {Xu}}]{Li:2019nux}%
  \BibitemOpen
  \bibfield  {author} {\bibinfo {author} {\bibfnamefont {E.-K.}\ \bibnamefont
  {Li}}, \bibinfo {author} {\bibfnamefont {M.}~\bibnamefont {Du}}, \bibinfo
  {author} {\bibfnamefont {Z.-H.}\ \bibnamefont {Zhou}}, \bibinfo {author}
  {\bibfnamefont {H.}~\bibnamefont {Zhang}}, \ and\ \bibinfo {author}
  {\bibfnamefont {L.}~\bibnamefont {Xu}},\ }\href {\doibase
  10.1093/mnras/staa3894} {\bibfield  {journal} {\bibinfo  {journal} {Mon. Not.
  Roy. Astron. Soc.}\ }\textbf {\bibinfo {volume} {501}},\ \bibinfo {pages}
  {4452} (\bibinfo {year} {2021})},\ \Eprint {http://arxiv.org/abs/1911.12076}
  {arXiv:1911.12076 [astro-ph.CO]} \BibitemShut {NoStop}%
\bibitem [{\citenamefont {{Stern}}\ \emph {et~al.}(2010)\citenamefont
  {{Stern}}, \citenamefont {{Jimenez}}, \citenamefont {{Verde}}, \citenamefont
  {{Kamionkowski}},\ and\ \citenamefont {{Stanford}}}]{2010JCAP...02..008S}%
  \BibitemOpen
  \bibfield  {author} {\bibinfo {author} {\bibfnamefont {D.}~\bibnamefont
  {{Stern}}}, \bibinfo {author} {\bibfnamefont {R.}~\bibnamefont {{Jimenez}}},
  \bibinfo {author} {\bibfnamefont {L.}~\bibnamefont {{Verde}}}, \bibinfo
  {author} {\bibfnamefont {M.}~\bibnamefont {{Kamionkowski}}}, \ and\ \bibinfo
  {author} {\bibfnamefont {S.~A.}\ \bibnamefont {{Stanford}}},\ }\href
  {\doibase 10.1088/1475-7516/2010/02/008} {\bibfield  {journal} {\bibinfo
  {journal} {jcap}\ }\textbf {\bibinfo {volume} {2010}},\ \bibinfo {eid} {008}
  (\bibinfo {year} {2010})},\ \Eprint {http://arxiv.org/abs/0907.3149}
  {arXiv:0907.3149 [astro-ph.CO]} \BibitemShut {NoStop}%
\bibitem [{\citenamefont {{Moresco}}\ \emph {et~al.}(2012)\citenamefont
  {{Moresco}}, \citenamefont {{Cimatti}}, \citenamefont {{Jimenez}},
  \citenamefont {{Pozzetti}}, \citenamefont {{Zamorani}}, \citenamefont
  {{Bolzonella}}, \citenamefont {{Dunlop}}, \citenamefont {{Lamareille}},
  \citenamefont {{Mignoli}}, \citenamefont {{Pearce}}, \citenamefont
  {{Rosati}}, \citenamefont {{Stern}}, \citenamefont {{Verde}}, \citenamefont
  {{Zucca}}, \citenamefont {{Carollo}}, \citenamefont {{Contini}},
  \citenamefont {{Kneib}}, \citenamefont {{Le F{\`e}vre}}, \citenamefont
  {{Lilly}}, \citenamefont {{Mainieri}}, \citenamefont {{Renzini}},
  \citenamefont {{Scodeggio}}, \citenamefont {{Balestra}}, \citenamefont
  {{Gobat}}, \citenamefont {{McLure}}, \citenamefont {{Bardelli}},
  \citenamefont {{Bongiorno}}, \citenamefont {{Caputi}}, \citenamefont
  {{Cucciati}}, \citenamefont {{de la Torre}}, \citenamefont {{de Ravel}},
  \citenamefont {{Franzetti}}, \citenamefont {{Garilli}}, \citenamefont
  {{Iovino}}, \citenamefont {{Kampczyk}}, \citenamefont {{Knobel}},
  \citenamefont {{Kova{\v{c}}}}, \citenamefont {{Le Borgne}}, \citenamefont
  {{Le Brun}}, \citenamefont {{Maier}}, \citenamefont {{Pell{\'o}}},
  \citenamefont {{Peng}}, \citenamefont {{Perez-Montero}}, \citenamefont
  {{Presotto}}, \citenamefont {{Silverman}}, \citenamefont {{Tanaka}},
  \citenamefont {{Tasca}}, \citenamefont {{Tresse}}, \citenamefont {{Vergani}},
  \citenamefont {{Almaini}}, \citenamefont {{Barnes}}, \citenamefont
  {{Bordoloi}}, \citenamefont {{Bradshaw}}, \citenamefont {{Cappi}},
  \citenamefont {{Chuter}}, \citenamefont {{Cirasuolo}}, \citenamefont
  {{Coppa}}, \citenamefont {{Diener}}, \citenamefont {{Foucaud}}, \citenamefont
  {{Hartley}}, \citenamefont {{Kamionkowski}}, \citenamefont {{Koekemoer}},
  \citenamefont {{L{\'o}pez-Sanjuan}}, \citenamefont {{McCracken}},
  \citenamefont {{Nair}}, \citenamefont {{Oesch}}, \citenamefont {{Stanford}},\
  and\ \citenamefont {{Welikala}}}]{2012JCAP...08..006M}%
  \BibitemOpen
  \bibfield  {author} {\bibinfo {author} {\bibfnamefont {M.}~\bibnamefont
  {{Moresco}}}, \bibinfo {author} {\bibfnamefont {A.}~\bibnamefont
  {{Cimatti}}}, \bibinfo {author} {\bibfnamefont {R.}~\bibnamefont
  {{Jimenez}}}, \bibinfo {author} {\bibfnamefont {L.}~\bibnamefont
  {{Pozzetti}}}, \bibinfo {author} {\bibfnamefont {G.}~\bibnamefont
  {{Zamorani}}}, \bibinfo {author} {\bibfnamefont {M.}~\bibnamefont
  {{Bolzonella}}}, \bibinfo {author} {\bibfnamefont {J.}~\bibnamefont
  {{Dunlop}}}, \bibinfo {author} {\bibfnamefont {F.}~\bibnamefont
  {{Lamareille}}}, \bibinfo {author} {\bibfnamefont {M.}~\bibnamefont
  {{Mignoli}}}, \bibinfo {author} {\bibfnamefont {H.}~\bibnamefont {{Pearce}}},
  \bibinfo {author} {\bibfnamefont {P.}~\bibnamefont {{Rosati}}}, \bibinfo
  {author} {\bibfnamefont {D.}~\bibnamefont {{Stern}}}, \bibinfo {author}
  {\bibfnamefont {L.}~\bibnamefont {{Verde}}}, \bibinfo {author} {\bibfnamefont
  {E.}~\bibnamefont {{Zucca}}}, \bibinfo {author} {\bibfnamefont {C.~M.}\
  \bibnamefont {{Carollo}}}, \bibinfo {author} {\bibfnamefont {T.}~\bibnamefont
  {{Contini}}}, \bibinfo {author} {\bibfnamefont {J.~P.}\ \bibnamefont
  {{Kneib}}}, \bibinfo {author} {\bibfnamefont {O.}~\bibnamefont {{Le
  F{\`e}vre}}}, \bibinfo {author} {\bibfnamefont {S.~J.}\ \bibnamefont
  {{Lilly}}}, \bibinfo {author} {\bibfnamefont {V.}~\bibnamefont {{Mainieri}}},
  \bibinfo {author} {\bibfnamefont {A.}~\bibnamefont {{Renzini}}}, \bibinfo
  {author} {\bibfnamefont {M.}~\bibnamefont {{Scodeggio}}}, \bibinfo {author}
  {\bibfnamefont {I.}~\bibnamefont {{Balestra}}}, \bibinfo {author}
  {\bibfnamefont {R.}~\bibnamefont {{Gobat}}}, \bibinfo {author} {\bibfnamefont
  {R.}~\bibnamefont {{McLure}}}, \bibinfo {author} {\bibfnamefont
  {S.}~\bibnamefont {{Bardelli}}}, \bibinfo {author} {\bibfnamefont
  {A.}~\bibnamefont {{Bongiorno}}}, \bibinfo {author} {\bibfnamefont
  {K.}~\bibnamefont {{Caputi}}}, \bibinfo {author} {\bibfnamefont
  {O.}~\bibnamefont {{Cucciati}}}, \bibinfo {author} {\bibfnamefont
  {S.}~\bibnamefont {{de la Torre}}}, \bibinfo {author} {\bibfnamefont
  {L.}~\bibnamefont {{de Ravel}}}, \bibinfo {author} {\bibfnamefont
  {P.}~\bibnamefont {{Franzetti}}}, \bibinfo {author} {\bibfnamefont
  {B.}~\bibnamefont {{Garilli}}}, \bibinfo {author} {\bibfnamefont
  {A.}~\bibnamefont {{Iovino}}}, \bibinfo {author} {\bibfnamefont
  {P.}~\bibnamefont {{Kampczyk}}}, \bibinfo {author} {\bibfnamefont
  {C.}~\bibnamefont {{Knobel}}}, \bibinfo {author} {\bibfnamefont
  {K.}~\bibnamefont {{Kova{\v{c}}}}}, \bibinfo {author} {\bibfnamefont {J.~F.}\
  \bibnamefont {{Le Borgne}}}, \bibinfo {author} {\bibfnamefont
  {V.}~\bibnamefont {{Le Brun}}}, \bibinfo {author} {\bibfnamefont
  {C.}~\bibnamefont {{Maier}}}, \bibinfo {author} {\bibfnamefont
  {R.}~\bibnamefont {{Pell{\'o}}}}, \bibinfo {author} {\bibfnamefont
  {Y.}~\bibnamefont {{Peng}}}, \bibinfo {author} {\bibfnamefont
  {E.}~\bibnamefont {{Perez-Montero}}}, \bibinfo {author} {\bibfnamefont
  {V.}~\bibnamefont {{Presotto}}}, \bibinfo {author} {\bibfnamefont {J.~D.}\
  \bibnamefont {{Silverman}}}, \bibinfo {author} {\bibfnamefont
  {M.}~\bibnamefont {{Tanaka}}}, \bibinfo {author} {\bibfnamefont {L.~A.~M.}\
  \bibnamefont {{Tasca}}}, \bibinfo {author} {\bibfnamefont {L.}~\bibnamefont
  {{Tresse}}}, \bibinfo {author} {\bibfnamefont {D.}~\bibnamefont {{Vergani}}},
  \bibinfo {author} {\bibfnamefont {O.}~\bibnamefont {{Almaini}}}, \bibinfo
  {author} {\bibfnamefont {L.}~\bibnamefont {{Barnes}}}, \bibinfo {author}
  {\bibfnamefont {R.}~\bibnamefont {{Bordoloi}}}, \bibinfo {author}
  {\bibfnamefont {E.}~\bibnamefont {{Bradshaw}}}, \bibinfo {author}
  {\bibfnamefont {A.}~\bibnamefont {{Cappi}}}, \bibinfo {author} {\bibfnamefont
  {R.}~\bibnamefont {{Chuter}}}, \bibinfo {author} {\bibfnamefont
  {M.}~\bibnamefont {{Cirasuolo}}}, \bibinfo {author} {\bibfnamefont
  {G.}~\bibnamefont {{Coppa}}}, \bibinfo {author} {\bibfnamefont
  {C.}~\bibnamefont {{Diener}}}, \bibinfo {author} {\bibfnamefont
  {S.}~\bibnamefont {{Foucaud}}}, \bibinfo {author} {\bibfnamefont
  {W.}~\bibnamefont {{Hartley}}}, \bibinfo {author} {\bibfnamefont
  {M.}~\bibnamefont {{Kamionkowski}}}, \bibinfo {author} {\bibfnamefont
  {A.~M.}\ \bibnamefont {{Koekemoer}}}, \bibinfo {author} {\bibfnamefont
  {C.}~\bibnamefont {{L{\'o}pez-Sanjuan}}}, \bibinfo {author} {\bibfnamefont
  {H.~J.}\ \bibnamefont {{McCracken}}}, \bibinfo {author} {\bibfnamefont
  {P.}~\bibnamefont {{Nair}}}, \bibinfo {author} {\bibfnamefont
  {P.}~\bibnamefont {{Oesch}}}, \bibinfo {author} {\bibfnamefont
  {A.}~\bibnamefont {{Stanford}}}, \ and\ \bibinfo {author} {\bibfnamefont
  {N.}~\bibnamefont {{Welikala}}},\ }\href {\doibase
  10.1088/1475-7516/2012/08/006} {\bibfield  {journal} {\bibinfo  {journal}
  {jcap}\ }\textbf {\bibinfo {volume} {2012}},\ \bibinfo {eid} {006} (\bibinfo
  {year} {2012})},\ \Eprint {http://arxiv.org/abs/1201.3609} {arXiv:1201.3609
  [astro-ph.CO]} \BibitemShut {NoStop}%
\bibitem [{\citenamefont {Moresco}\ \emph {et~al.}(2016)\citenamefont
  {Moresco}, \citenamefont {Pozzetti}, \citenamefont {Cimatti}, \citenamefont
  {Jimenez}, \citenamefont {Maraston}, \citenamefont {Verde}, \citenamefont
  {Thomas}, \citenamefont {Citro}, \citenamefont {Tojeiro},\ and\ \citenamefont
  {Wilkinson}}]{Moresco:2016mzx}%
  \BibitemOpen
  \bibfield  {author} {\bibinfo {author} {\bibfnamefont {M.}~\bibnamefont
  {Moresco}}, \bibinfo {author} {\bibfnamefont {L.}~\bibnamefont {Pozzetti}},
  \bibinfo {author} {\bibfnamefont {A.}~\bibnamefont {Cimatti}}, \bibinfo
  {author} {\bibfnamefont {R.}~\bibnamefont {Jimenez}}, \bibinfo {author}
  {\bibfnamefont {C.}~\bibnamefont {Maraston}}, \bibinfo {author}
  {\bibfnamefont {L.}~\bibnamefont {Verde}}, \bibinfo {author} {\bibfnamefont
  {D.}~\bibnamefont {Thomas}}, \bibinfo {author} {\bibfnamefont
  {A.}~\bibnamefont {Citro}}, \bibinfo {author} {\bibfnamefont
  {R.}~\bibnamefont {Tojeiro}}, \ and\ \bibinfo {author} {\bibfnamefont
  {D.}~\bibnamefont {Wilkinson}},\ }\href {\doibase
  10.1088/1475-7516/2016/05/014} {\bibfield  {journal} {\bibinfo  {journal}
  {JCAP}\ }\textbf {\bibinfo {volume} {05}},\ \bibinfo {pages} {014} (\bibinfo
  {year} {2016})},\ \Eprint {http://arxiv.org/abs/1601.01701} {arXiv:1601.01701
  [astro-ph.CO]} \BibitemShut {NoStop}%
\bibitem [{\citenamefont {Moresco}(2015)}]{Moresco:2015cya}%
  \BibitemOpen
  \bibfield  {author} {\bibinfo {author} {\bibfnamefont {M.}~\bibnamefont
  {Moresco}},\ }\href {\doibase 10.1093/mnrasl/slv037} {\bibfield  {journal}
  {\bibinfo  {journal} {Mon. Not. Roy. Astron. Soc.}\ }\textbf {\bibinfo
  {volume} {450}},\ \bibinfo {pages} {L16} (\bibinfo {year} {2015})},\ \Eprint
  {http://arxiv.org/abs/1503.01116} {arXiv:1503.01116 [astro-ph.CO]}
  \BibitemShut {NoStop}%
\bibitem [{\citenamefont {Ratsimbazafy}\ \emph {et~al.}(2017)\citenamefont
  {Ratsimbazafy}, \citenamefont {Loubser}, \citenamefont {Crawford},
  \citenamefont {Cress}, \citenamefont {Bassett}, \citenamefont {Nichol},\ and\
  \citenamefont {V\"ais\"anen}}]{Ratsimbazafy:2017vga}%
  \BibitemOpen
  \bibfield  {author} {\bibinfo {author} {\bibfnamefont {A.~L.}\ \bibnamefont
  {Ratsimbazafy}}, \bibinfo {author} {\bibfnamefont {S.~I.}\ \bibnamefont
  {Loubser}}, \bibinfo {author} {\bibfnamefont {S.~M.}\ \bibnamefont
  {Crawford}}, \bibinfo {author} {\bibfnamefont {C.~M.}\ \bibnamefont {Cress}},
  \bibinfo {author} {\bibfnamefont {B.~A.}\ \bibnamefont {Bassett}}, \bibinfo
  {author} {\bibfnamefont {R.~C.}\ \bibnamefont {Nichol}}, \ and\ \bibinfo
  {author} {\bibfnamefont {P.}~\bibnamefont {V\"ais\"anen}},\ }\href {\doibase
  10.1093/mnras/stx301} {\bibfield  {journal} {\bibinfo  {journal} {Mon. Not.
  Roy. Astron. Soc.}\ }\textbf {\bibinfo {volume} {467}},\ \bibinfo {pages}
  {3239} (\bibinfo {year} {2017})},\ \Eprint {http://arxiv.org/abs/1702.00418}
  {arXiv:1702.00418 [astro-ph.CO]} \BibitemShut {NoStop}%
\bibitem [{\citenamefont {{Zhang}}\ \emph {et~al.}(2014)\citenamefont
  {{Zhang}}, \citenamefont {{Zhang}}, \citenamefont {{Yuan}}, \citenamefont
  {{Liu}}, \citenamefont {{Zhang}},\ and\ \citenamefont
  {{Sun}}}]{2014RAA....14.1221Z}%
  \BibitemOpen
  \bibfield  {author} {\bibinfo {author} {\bibfnamefont {C.}~\bibnamefont
  {{Zhang}}}, \bibinfo {author} {\bibfnamefont {H.}~\bibnamefont {{Zhang}}},
  \bibinfo {author} {\bibfnamefont {S.}~\bibnamefont {{Yuan}}}, \bibinfo
  {author} {\bibfnamefont {S.}~\bibnamefont {{Liu}}}, \bibinfo {author}
  {\bibfnamefont {T.-J.}\ \bibnamefont {{Zhang}}}, \ and\ \bibinfo {author}
  {\bibfnamefont {Y.-C.}\ \bibnamefont {{Sun}}},\ }\href {\doibase
  10.1088/1674-4527/14/10/002} {\bibfield  {journal} {\bibinfo  {journal}
  {Research in Astronomy and Astrophysics}\ }\textbf {\bibinfo {volume} {14}},\
  \bibinfo {eid} {1221-1233} (\bibinfo {year} {2014})},\ \Eprint
  {http://arxiv.org/abs/1207.4541} {arXiv:1207.4541 [astro-ph.CO]} \BibitemShut
  {NoStop}%
\bibitem [{\citenamefont {Singirikonda}\ and\ \citenamefont
  {Desai}(2020)}]{Singirikonda:2020ieg}%
  \BibitemOpen
  \bibfield  {author} {\bibinfo {author} {\bibfnamefont {H.}~\bibnamefont
  {Singirikonda}}\ and\ \bibinfo {author} {\bibfnamefont {S.}~\bibnamefont
  {Desai}},\ }\href {\doibase 10.1140/epjc/s10052-020-8289-8} {\bibfield
  {journal} {\bibinfo  {journal} {Eur. Phys. J. C}\ }\textbf {\bibinfo {volume}
  {80}},\ \bibinfo {pages} {694} (\bibinfo {year} {2020})},\ \Eprint
  {http://arxiv.org/abs/2003.00494} {arXiv:2003.00494 [astro-ph.CO]}
  \BibitemShut {NoStop}%
\bibitem [{\citenamefont {Scolnic}\ \emph {et~al.}(2018)\citenamefont {Scolnic}
  \emph {et~al.}}]{Pan-STARRS1:2017jku}%
  \BibitemOpen
  \bibfield  {author} {\bibinfo {author} {\bibfnamefont {D.~M.}\ \bibnamefont
  {Scolnic}} \emph {et~al.} (\bibinfo {collaboration} {Pan-STARRS1}),\ }\href
  {\doibase 10.3847/1538-4357/aab9bb} {\bibfield  {journal} {\bibinfo
  {journal} {Astrophys. J.}\ }\textbf {\bibinfo {volume} {859}},\ \bibinfo
  {pages} {101} (\bibinfo {year} {2018})},\ \Eprint
  {http://arxiv.org/abs/1710.00845} {arXiv:1710.00845 [astro-ph.CO]}
  \BibitemShut {NoStop}%
\bibitem [{\citenamefont {Song}\ and\ \citenamefont
  {Percival}(2009)}]{Song:2008qt}%
  \BibitemOpen
  \bibfield  {author} {\bibinfo {author} {\bibfnamefont {Y.-S.}\ \bibnamefont
  {Song}}\ and\ \bibinfo {author} {\bibfnamefont {W.~J.}\ \bibnamefont
  {Percival}},\ }\href {\doibase 10.1088/1475-7516/2009/10/004} {\bibfield
  {journal} {\bibinfo  {journal} {JCAP}\ }\textbf {\bibinfo {volume} {10}},\
  \bibinfo {pages} {004} (\bibinfo {year} {2009})},\ \Eprint
  {http://arxiv.org/abs/0807.0810} {arXiv:0807.0810 [astro-ph]} \BibitemShut
  {NoStop}%
\bibitem [{\citenamefont {Davis}\ \emph {et~al.}(2011)\citenamefont {Davis},
  \citenamefont {Nusser}, \citenamefont {Masters}, \citenamefont {Springob},
  \citenamefont {Huchra},\ and\ \citenamefont {Lemson}}]{Davis:2010sw}%
  \BibitemOpen
  \bibfield  {author} {\bibinfo {author} {\bibfnamefont {M.}~\bibnamefont
  {Davis}}, \bibinfo {author} {\bibfnamefont {A.}~\bibnamefont {Nusser}},
  \bibinfo {author} {\bibfnamefont {K.}~\bibnamefont {Masters}}, \bibinfo
  {author} {\bibfnamefont {C.}~\bibnamefont {Springob}}, \bibinfo {author}
  {\bibfnamefont {J.~P.}\ \bibnamefont {Huchra}}, \ and\ \bibinfo {author}
  {\bibfnamefont {G.}~\bibnamefont {Lemson}},\ }\href {\doibase
  10.1111/j.1365-2966.2011.18362.x} {\bibfield  {journal} {\bibinfo  {journal}
  {Mon. Not. Roy. Astron. Soc.}\ }\textbf {\bibinfo {volume} {413}},\ \bibinfo
  {pages} {2906} (\bibinfo {year} {2011})},\ \Eprint
  {http://arxiv.org/abs/1011.3114} {arXiv:1011.3114 [astro-ph.CO]} \BibitemShut
  {NoStop}%
\bibitem [{\citenamefont {Hudson}\ and\ \citenamefont
  {Turnbull}(2012)}]{Hudson_2012}%
  \BibitemOpen
  \bibfield  {author} {\bibinfo {author} {\bibfnamefont {M.~J.}\ \bibnamefont
  {Hudson}}\ and\ \bibinfo {author} {\bibfnamefont {S.~J.}\ \bibnamefont
  {Turnbull}},\ }\href {\doibase 10.1088/2041-8205/751/2/l30} {\bibfield
  {journal} {\bibinfo  {journal} {The Astrophysical Journal}\ }\textbf
  {\bibinfo {volume} {751}},\ \bibinfo {pages} {L30} (\bibinfo {year}
  {2012})}\BibitemShut {NoStop}%
\bibitem [{\citenamefont {Turnbull}\ \emph {et~al.}(2011)\citenamefont
  {Turnbull}, \citenamefont {Hudson}, \citenamefont {Feldman}, \citenamefont
  {Hicken}, \citenamefont {Kirshner},\ and\ \citenamefont
  {Watkins}}]{Turnbull_2011}%
  \BibitemOpen
  \bibfield  {author} {\bibinfo {author} {\bibfnamefont {S.~J.}\ \bibnamefont
  {Turnbull}}, \bibinfo {author} {\bibfnamefont {M.~J.}\ \bibnamefont
  {Hudson}}, \bibinfo {author} {\bibfnamefont {H.~A.}\ \bibnamefont {Feldman}},
  \bibinfo {author} {\bibfnamefont {M.}~\bibnamefont {Hicken}}, \bibinfo
  {author} {\bibfnamefont {R.~P.}\ \bibnamefont {Kirshner}}, \ and\ \bibinfo
  {author} {\bibfnamefont {R.}~\bibnamefont {Watkins}},\ }\href {\doibase
  10.1111/j.1365-2966.2011.20050.x} {\bibfield  {journal} {\bibinfo  {journal}
  {Monthly Notices of the Royal Astronomical Society}\ }\textbf {\bibinfo
  {volume} {420}},\ \bibinfo {pages} {447–454} (\bibinfo {year}
  {2011})}\BibitemShut {NoStop}%
\bibitem [{\citenamefont {Samushia}\ \emph {et~al.}(2012)\citenamefont
  {Samushia}, \citenamefont {Percival},\ and\ \citenamefont
  {Raccanelli}}]{Samushia_2012}%
  \BibitemOpen
  \bibfield  {author} {\bibinfo {author} {\bibfnamefont {L.}~\bibnamefont
  {Samushia}}, \bibinfo {author} {\bibfnamefont {W.~J.}\ \bibnamefont
  {Percival}}, \ and\ \bibinfo {author} {\bibfnamefont {A.}~\bibnamefont
  {Raccanelli}},\ }\href {\doibase 10.1111/j.1365-2966.2011.20169.x} {\bibfield
   {journal} {\bibinfo  {journal} {Monthly Notices of the Royal Astronomical
  Society}\ }\textbf {\bibinfo {volume} {420}},\ \bibinfo {pages} {2102–2119}
  (\bibinfo {year} {2012})}\BibitemShut {NoStop}%
\bibitem [{\citenamefont {Blake}\ \emph
  {et~al.}(2012{\natexlab{a}})\citenamefont {Blake}, \citenamefont {Brough},
  \citenamefont {Colless}, \citenamefont {Contreras}, \citenamefont {Couch},
  \citenamefont {Croom}, \citenamefont {Croton}, \citenamefont {Davis},
  \citenamefont {Drinkwater}, \citenamefont {Forster}, \citenamefont {Gilbank},
  \citenamefont {Gladders}, \citenamefont {Glazebrook}, \citenamefont
  {Jelliffe}, \citenamefont {Jurek}, \citenamefont {Li}, \citenamefont
  {Madore}, \citenamefont {Martin}, \citenamefont {Pimbblet}, \citenamefont
  {Poole}, \citenamefont {Pracy}, \citenamefont {Sharp}, \citenamefont
  {Wisnioski}, \citenamefont {Woods}, \citenamefont {Wyder},\ and\
  \citenamefont {Yee}}]{Blake_2012}%
  \BibitemOpen
  \bibfield  {author} {\bibinfo {author} {\bibfnamefont {C.}~\bibnamefont
  {Blake}}, \bibinfo {author} {\bibfnamefont {S.}~\bibnamefont {Brough}},
  \bibinfo {author} {\bibfnamefont {M.}~\bibnamefont {Colless}}, \bibinfo
  {author} {\bibfnamefont {C.}~\bibnamefont {Contreras}}, \bibinfo {author}
  {\bibfnamefont {W.}~\bibnamefont {Couch}}, \bibinfo {author} {\bibfnamefont
  {S.}~\bibnamefont {Croom}}, \bibinfo {author} {\bibfnamefont
  {D.}~\bibnamefont {Croton}}, \bibinfo {author} {\bibfnamefont {T.~M.}\
  \bibnamefont {Davis}}, \bibinfo {author} {\bibfnamefont {M.~J.}\ \bibnamefont
  {Drinkwater}}, \bibinfo {author} {\bibfnamefont {K.}~\bibnamefont {Forster}},
  \bibinfo {author} {\bibfnamefont {D.}~\bibnamefont {Gilbank}}, \bibinfo
  {author} {\bibfnamefont {M.}~\bibnamefont {Gladders}}, \bibinfo {author}
  {\bibfnamefont {K.}~\bibnamefont {Glazebrook}}, \bibinfo {author}
  {\bibfnamefont {B.}~\bibnamefont {Jelliffe}}, \bibinfo {author}
  {\bibfnamefont {R.~J.}\ \bibnamefont {Jurek}}, \bibinfo {author}
  {\bibfnamefont {I.-h.}\ \bibnamefont {Li}}, \bibinfo {author} {\bibfnamefont
  {B.}~\bibnamefont {Madore}}, \bibinfo {author} {\bibfnamefont {D.~C.}\
  \bibnamefont {Martin}}, \bibinfo {author} {\bibfnamefont {K.}~\bibnamefont
  {Pimbblet}}, \bibinfo {author} {\bibfnamefont {G.~B.}\ \bibnamefont {Poole}},
  \bibinfo {author} {\bibfnamefont {M.}~\bibnamefont {Pracy}}, \bibinfo
  {author} {\bibfnamefont {R.}~\bibnamefont {Sharp}}, \bibinfo {author}
  {\bibfnamefont {E.}~\bibnamefont {Wisnioski}}, \bibinfo {author}
  {\bibfnamefont {D.}~\bibnamefont {Woods}}, \bibinfo {author} {\bibfnamefont
  {T.~K.}\ \bibnamefont {Wyder}}, \ and\ \bibinfo {author} {\bibfnamefont
  {H.~K.~C.}\ \bibnamefont {Yee}},\ }\href {\doibase
  10.1111/j.1365-2966.2012.21473.x} {\bibfield  {journal} {\bibinfo  {journal}
  {Monthly Notices of the Royal Astronomical Society}\ }\textbf {\bibinfo
  {volume} {425}},\ \bibinfo {pages} {405–414} (\bibinfo {year}
  {2012}{\natexlab{a}})}\BibitemShut {NoStop}%
\bibitem [{\citenamefont {Tojeiro}\ \emph {et~al.}(2012)\citenamefont
  {Tojeiro}, \citenamefont {Percival}, \citenamefont {Brinkmann}, \citenamefont
  {Brownstein}, \citenamefont {Eisenstein}, \citenamefont {Manera},
  \citenamefont {Maraston}, \citenamefont {McBride}, \citenamefont {Muna},
  \citenamefont {Reid}, \citenamefont {Ross}, \citenamefont {Ross},
  \citenamefont {Samushia}, \citenamefont {Padmanabhan}, \citenamefont
  {Schneider}, \citenamefont {Skibba}, \citenamefont {Sánchez}, \citenamefont
  {Swanson}, \citenamefont {Thomas}, \citenamefont {Tinker}, \citenamefont
  {Verde}, \citenamefont {Wake}, \citenamefont {Weaver},\ and\ \citenamefont
  {Zhao}}]{Tojeiro_2012}%
  \BibitemOpen
  \bibfield  {author} {\bibinfo {author} {\bibfnamefont {R.}~\bibnamefont
  {Tojeiro}}, \bibinfo {author} {\bibfnamefont {W.~J.}\ \bibnamefont
  {Percival}}, \bibinfo {author} {\bibfnamefont {J.}~\bibnamefont {Brinkmann}},
  \bibinfo {author} {\bibfnamefont {J.~R.}\ \bibnamefont {Brownstein}},
  \bibinfo {author} {\bibfnamefont {D.~J.}\ \bibnamefont {Eisenstein}},
  \bibinfo {author} {\bibfnamefont {M.}~\bibnamefont {Manera}}, \bibinfo
  {author} {\bibfnamefont {C.}~\bibnamefont {Maraston}}, \bibinfo {author}
  {\bibfnamefont {C.~K.}\ \bibnamefont {McBride}}, \bibinfo {author}
  {\bibfnamefont {D.}~\bibnamefont {Muna}}, \bibinfo {author} {\bibfnamefont
  {B.}~\bibnamefont {Reid}}, \bibinfo {author} {\bibfnamefont {A.~J.}\
  \bibnamefont {Ross}}, \bibinfo {author} {\bibfnamefont {N.~P.}\ \bibnamefont
  {Ross}}, \bibinfo {author} {\bibfnamefont {L.}~\bibnamefont {Samushia}},
  \bibinfo {author} {\bibfnamefont {N.}~\bibnamefont {Padmanabhan}}, \bibinfo
  {author} {\bibfnamefont {D.~P.}\ \bibnamefont {Schneider}}, \bibinfo {author}
  {\bibfnamefont {R.}~\bibnamefont {Skibba}}, \bibinfo {author} {\bibfnamefont
  {A.~G.}\ \bibnamefont {Sánchez}}, \bibinfo {author} {\bibfnamefont
  {M.~E.~C.}\ \bibnamefont {Swanson}}, \bibinfo {author} {\bibfnamefont
  {D.}~\bibnamefont {Thomas}}, \bibinfo {author} {\bibfnamefont {J.~L.}\
  \bibnamefont {Tinker}}, \bibinfo {author} {\bibfnamefont {L.}~\bibnamefont
  {Verde}}, \bibinfo {author} {\bibfnamefont {D.~A.}\ \bibnamefont {Wake}},
  \bibinfo {author} {\bibfnamefont {B.~A.}\ \bibnamefont {Weaver}}, \ and\
  \bibinfo {author} {\bibfnamefont {G.-B.}\ \bibnamefont {Zhao}},\ }\href
  {\doibase 10.1111/j.1365-2966.2012.21404.x} {\bibfield  {journal} {\bibinfo
  {journal} {Monthly Notices of the Royal Astronomical Society}\ }\textbf
  {\bibinfo {volume} {424}},\ \bibinfo {pages} {2339–2344} (\bibinfo {year}
  {2012})}\BibitemShut {NoStop}%
\bibitem [{\citenamefont {Chuang}\ and\ \citenamefont
  {Wang}(2013)}]{Chuang_2013}%
  \BibitemOpen
  \bibfield  {author} {\bibinfo {author} {\bibfnamefont {C.-H.}\ \bibnamefont
  {Chuang}}\ and\ \bibinfo {author} {\bibfnamefont {Y.}~\bibnamefont {Wang}},\
  }\href {\doibase 10.1093/mnras/stt1290} {\bibfield  {journal} {\bibinfo
  {journal} {Monthly Notices of the Royal Astronomical Society}\ }\textbf
  {\bibinfo {volume} {435}},\ \bibinfo {pages} {255–262} (\bibinfo {year}
  {2013})}\BibitemShut {NoStop}%
\bibitem [{\citenamefont {Beutler}\ \emph {et~al.}(2012)\citenamefont
  {Beutler}, \citenamefont {Blake}, \citenamefont {Colless}, \citenamefont
  {Jones}, \citenamefont {Staveley-Smith}, \citenamefont {Poole}, \citenamefont
  {Campbell}, \citenamefont {Parker}, \citenamefont {Saunders},\ and\
  \citenamefont {Watson}}]{Beutler_2012}%
  \BibitemOpen
  \bibfield  {author} {\bibinfo {author} {\bibfnamefont {F.}~\bibnamefont
  {Beutler}}, \bibinfo {author} {\bibfnamefont {C.}~\bibnamefont {Blake}},
  \bibinfo {author} {\bibfnamefont {M.}~\bibnamefont {Colless}}, \bibinfo
  {author} {\bibfnamefont {D.~H.}\ \bibnamefont {Jones}}, \bibinfo {author}
  {\bibfnamefont {L.}~\bibnamefont {Staveley-Smith}}, \bibinfo {author}
  {\bibfnamefont {G.~B.}\ \bibnamefont {Poole}}, \bibinfo {author}
  {\bibfnamefont {L.}~\bibnamefont {Campbell}}, \bibinfo {author}
  {\bibfnamefont {Q.}~\bibnamefont {Parker}}, \bibinfo {author} {\bibfnamefont
  {W.}~\bibnamefont {Saunders}}, \ and\ \bibinfo {author} {\bibfnamefont
  {F.}~\bibnamefont {Watson}},\ }\href {\doibase
  10.1111/j.1365-2966.2012.21136.x} {\bibfield  {journal} {\bibinfo  {journal}
  {Monthly Notices of the Royal Astronomical Society}\ }\textbf {\bibinfo
  {volume} {423}},\ \bibinfo {pages} {3430–3444} (\bibinfo {year}
  {2012})}\BibitemShut {NoStop}%
\bibitem [{\citenamefont {Blake}\ \emph {et~al.}(2013)\citenamefont {Blake},
  \citenamefont {Baldry}, \citenamefont {Bland-Hawthorn}, \citenamefont
  {Christodoulou}, \citenamefont {Colless}, \citenamefont {Conselice},
  \citenamefont {Driver}, \citenamefont {Hopkins}, \citenamefont {Liske},
  \citenamefont {Loveday}, \citenamefont {Norberg}, \citenamefont {Peacock},
  \citenamefont {Poole},\ and\ \citenamefont {Robotham}}]{Blake_2013}%
  \BibitemOpen
  \bibfield  {author} {\bibinfo {author} {\bibfnamefont {C.}~\bibnamefont
  {Blake}}, \bibinfo {author} {\bibfnamefont {I.~K.}\ \bibnamefont {Baldry}},
  \bibinfo {author} {\bibfnamefont {J.}~\bibnamefont {Bland-Hawthorn}},
  \bibinfo {author} {\bibfnamefont {L.}~\bibnamefont {Christodoulou}}, \bibinfo
  {author} {\bibfnamefont {M.}~\bibnamefont {Colless}}, \bibinfo {author}
  {\bibfnamefont {C.}~\bibnamefont {Conselice}}, \bibinfo {author}
  {\bibfnamefont {S.~P.}\ \bibnamefont {Driver}}, \bibinfo {author}
  {\bibfnamefont {A.~M.}\ \bibnamefont {Hopkins}}, \bibinfo {author}
  {\bibfnamefont {J.}~\bibnamefont {Liske}}, \bibinfo {author} {\bibfnamefont
  {J.}~\bibnamefont {Loveday}}, \bibinfo {author} {\bibfnamefont
  {P.}~\bibnamefont {Norberg}}, \bibinfo {author} {\bibfnamefont {J.~A.}\
  \bibnamefont {Peacock}}, \bibinfo {author} {\bibfnamefont {G.~B.}\
  \bibnamefont {Poole}}, \ and\ \bibinfo {author} {\bibfnamefont {A.~S.~G.}\
  \bibnamefont {Robotham}},\ }\href {\doibase 10.1093/mnras/stt1791} {\bibfield
   {journal} {\bibinfo  {journal} {Monthly Notices of the Royal Astronomical
  Society}\ }\textbf {\bibinfo {volume} {436}},\ \bibinfo {pages} {3089–3105}
  (\bibinfo {year} {2013})}\BibitemShut {NoStop}%
\bibitem [{\citenamefont {Sánchez}\ \emph {et~al.}(2014)\citenamefont
  {Sánchez}, \citenamefont {Montesano}, \citenamefont {Kazin}, \citenamefont
  {Aubourg}, \citenamefont {Beutler}, \citenamefont {Brinkmann}, \citenamefont
  {Brownstein}, \citenamefont {Cuesta}, \citenamefont {Dawson}, \citenamefont
  {Eisenstein}, \citenamefont {Ho}, \citenamefont {Honscheid}, \citenamefont
  {Manera}, \citenamefont {Maraston}, \citenamefont {McBride}, \citenamefont
  {Percival}, \citenamefont {Ross}, \citenamefont {Samushia}, \citenamefont
  {Schlegel}, \citenamefont {Schneider}, \citenamefont {Skibba}, \citenamefont
  {Thomas}, \citenamefont {Tinker}, \citenamefont {Tojeiro}, \citenamefont
  {Wake}, \citenamefont {Weaver}, \citenamefont {White},\ and\ \citenamefont
  {Zehavi}}]{S_nchez_2014}%
  \BibitemOpen
  \bibfield  {author} {\bibinfo {author} {\bibfnamefont {A.~G.}\ \bibnamefont
  {Sánchez}}, \bibinfo {author} {\bibfnamefont {F.}~\bibnamefont {Montesano}},
  \bibinfo {author} {\bibfnamefont {E.~A.}\ \bibnamefont {Kazin}}, \bibinfo
  {author} {\bibfnamefont {E.}~\bibnamefont {Aubourg}}, \bibinfo {author}
  {\bibfnamefont {F.}~\bibnamefont {Beutler}}, \bibinfo {author} {\bibfnamefont
  {J.}~\bibnamefont {Brinkmann}}, \bibinfo {author} {\bibfnamefont {J.~R.}\
  \bibnamefont {Brownstein}}, \bibinfo {author} {\bibfnamefont {A.~J.}\
  \bibnamefont {Cuesta}}, \bibinfo {author} {\bibfnamefont {K.~S.}\
  \bibnamefont {Dawson}}, \bibinfo {author} {\bibfnamefont {D.~J.}\
  \bibnamefont {Eisenstein}}, \bibinfo {author} {\bibfnamefont
  {S.}~\bibnamefont {Ho}}, \bibinfo {author} {\bibfnamefont {K.}~\bibnamefont
  {Honscheid}}, \bibinfo {author} {\bibfnamefont {M.}~\bibnamefont {Manera}},
  \bibinfo {author} {\bibfnamefont {C.}~\bibnamefont {Maraston}}, \bibinfo
  {author} {\bibfnamefont {C.~K.}\ \bibnamefont {McBride}}, \bibinfo {author}
  {\bibfnamefont {W.~J.}\ \bibnamefont {Percival}}, \bibinfo {author}
  {\bibfnamefont {A.~J.}\ \bibnamefont {Ross}}, \bibinfo {author}
  {\bibfnamefont {L.}~\bibnamefont {Samushia}}, \bibinfo {author}
  {\bibfnamefont {D.~J.}\ \bibnamefont {Schlegel}}, \bibinfo {author}
  {\bibfnamefont {D.~P.}\ \bibnamefont {Schneider}}, \bibinfo {author}
  {\bibfnamefont {R.}~\bibnamefont {Skibba}}, \bibinfo {author} {\bibfnamefont
  {D.}~\bibnamefont {Thomas}}, \bibinfo {author} {\bibfnamefont {J.~L.}\
  \bibnamefont {Tinker}}, \bibinfo {author} {\bibfnamefont {R.}~\bibnamefont
  {Tojeiro}}, \bibinfo {author} {\bibfnamefont {D.~A.}\ \bibnamefont {Wake}},
  \bibinfo {author} {\bibfnamefont {B.~A.}\ \bibnamefont {Weaver}}, \bibinfo
  {author} {\bibfnamefont {M.}~\bibnamefont {White}}, \ and\ \bibinfo {author}
  {\bibfnamefont {I.}~\bibnamefont {Zehavi}},\ }\href {\doibase
  10.1093/mnras/stu342} {\bibfield  {journal} {\bibinfo  {journal} {Monthly
  Notices of the Royal Astronomical Society}\ }\textbf {\bibinfo {volume}
  {440}},\ \bibinfo {pages} {2692–2713} (\bibinfo {year} {2014})}\BibitemShut
  {NoStop}%
\bibitem [{\citenamefont {Chuang}\ \emph {et~al.}(2016)\citenamefont {Chuang},
  \citenamefont {Prada}, \citenamefont {Pellejero-Ibanez}, \citenamefont
  {Beutler}, \citenamefont {Cuesta}, \citenamefont {Eisenstein}, \citenamefont
  {Escoffier}, \citenamefont {Ho}, \citenamefont {Kitaura}, \citenamefont
  {Kneib}, \citenamefont {Manera}, \citenamefont {Nuza}, \citenamefont
  {Rodríguez-Torres}, \citenamefont {Ross}, \citenamefont {Rubiño-Martín},
  \citenamefont {Samushia}, \citenamefont {Schlegel}, \citenamefont
  {Schneider}, \citenamefont {Wang}, \citenamefont {Weaver}, \citenamefont
  {Zhao}, \citenamefont {Brownstein}, \citenamefont {Dawson}, \citenamefont
  {Maraston}, \citenamefont {Olmstead},\ and\ \citenamefont
  {Thomas}}]{Chuang_2016}%
  \BibitemOpen
  \bibfield  {author} {\bibinfo {author} {\bibfnamefont {C.-H.}\ \bibnamefont
  {Chuang}}, \bibinfo {author} {\bibfnamefont {F.}~\bibnamefont {Prada}},
  \bibinfo {author} {\bibfnamefont {M.}~\bibnamefont {Pellejero-Ibanez}},
  \bibinfo {author} {\bibfnamefont {F.}~\bibnamefont {Beutler}}, \bibinfo
  {author} {\bibfnamefont {A.~J.}\ \bibnamefont {Cuesta}}, \bibinfo {author}
  {\bibfnamefont {D.~J.}\ \bibnamefont {Eisenstein}}, \bibinfo {author}
  {\bibfnamefont {S.}~\bibnamefont {Escoffier}}, \bibinfo {author}
  {\bibfnamefont {S.}~\bibnamefont {Ho}}, \bibinfo {author} {\bibfnamefont
  {F.-S.}\ \bibnamefont {Kitaura}}, \bibinfo {author} {\bibfnamefont {J.-P.}\
  \bibnamefont {Kneib}}, \bibinfo {author} {\bibfnamefont {M.}~\bibnamefont
  {Manera}}, \bibinfo {author} {\bibfnamefont {S.~E.}\ \bibnamefont {Nuza}},
  \bibinfo {author} {\bibfnamefont {S.}~\bibnamefont {Rodríguez-Torres}},
  \bibinfo {author} {\bibfnamefont {A.}~\bibnamefont {Ross}}, \bibinfo {author}
  {\bibfnamefont {J.~A.}\ \bibnamefont {Rubiño-Martín}}, \bibinfo {author}
  {\bibfnamefont {L.}~\bibnamefont {Samushia}}, \bibinfo {author}
  {\bibfnamefont {D.~J.}\ \bibnamefont {Schlegel}}, \bibinfo {author}
  {\bibfnamefont {D.~P.}\ \bibnamefont {Schneider}}, \bibinfo {author}
  {\bibfnamefont {Y.}~\bibnamefont {Wang}}, \bibinfo {author} {\bibfnamefont
  {B.~A.}\ \bibnamefont {Weaver}}, \bibinfo {author} {\bibfnamefont
  {G.}~\bibnamefont {Zhao}}, \bibinfo {author} {\bibfnamefont {J.~R.}\
  \bibnamefont {Brownstein}}, \bibinfo {author} {\bibfnamefont {K.~S.}\
  \bibnamefont {Dawson}}, \bibinfo {author} {\bibfnamefont {C.}~\bibnamefont
  {Maraston}}, \bibinfo {author} {\bibfnamefont {M.~D.}\ \bibnamefont
  {Olmstead}}, \ and\ \bibinfo {author} {\bibfnamefont {D.}~\bibnamefont
  {Thomas}},\ }\href {\doibase 10.1093/mnras/stw1535} {\bibfield  {journal}
  {\bibinfo  {journal} {Monthly Notices of the Royal Astronomical Society}\
  }\textbf {\bibinfo {volume} {461}},\ \bibinfo {pages} {3781–3793} (\bibinfo
  {year} {2016})}\BibitemShut {NoStop}%
\bibitem [{\citenamefont {de~la Torre}\ \emph {et~al.}(2013)\citenamefont
  {de~la Torre}, \citenamefont {Guzzo}, \citenamefont {Peacock}, \citenamefont
  {Branchini}, \citenamefont {Iovino}, \citenamefont {Granett}, \citenamefont
  {Abbas}, \citenamefont {Adami}, \citenamefont {Arnouts}, \citenamefont {Bel},
  \citenamefont {Bolzonella}, \citenamefont {Bottini}, \citenamefont {Cappi},
  \citenamefont {Coupon}, \citenamefont {Cucciati}, \citenamefont {Davidzon},
  \citenamefont {De~Lucia}, \citenamefont {Fritz}, \citenamefont {Franzetti},
  \citenamefont {Fumana}, \citenamefont {Garilli}, \citenamefont {Ilbert},
  \citenamefont {Krywult}, \citenamefont {Le~Brun}, \citenamefont {Le~Fèvre},
  \citenamefont {Maccagni}, \citenamefont {Małek}, \citenamefont {Marulli},
  \citenamefont {McCracken}, \citenamefont {Moscardini}, \citenamefont
  {Paioro}, \citenamefont {Percival}, \citenamefont {Polletta}, \citenamefont
  {Pollo}, \citenamefont {Schlagenhaufer}, \citenamefont {Scodeggio},
  \citenamefont {Tasca}, \citenamefont {Tojeiro}, \citenamefont {Vergani},
  \citenamefont {Zanichelli}, \citenamefont {Burden}, \citenamefont {Di~Porto},
  \citenamefont {Marchetti}, \citenamefont {Marinoni}, \citenamefont {Mellier},
  \citenamefont {Monaco}, \citenamefont {Nichol}, \citenamefont {Phleps},
  \citenamefont {Wolk},\ and\ \citenamefont {Zamorani}}]{de_la_Torre_2013}%
  \BibitemOpen
  \bibfield  {author} {\bibinfo {author} {\bibfnamefont {S.}~\bibnamefont
  {de~la Torre}}, \bibinfo {author} {\bibfnamefont {L.}~\bibnamefont {Guzzo}},
  \bibinfo {author} {\bibfnamefont {J.~A.}\ \bibnamefont {Peacock}}, \bibinfo
  {author} {\bibfnamefont {E.}~\bibnamefont {Branchini}}, \bibinfo {author}
  {\bibfnamefont {A.}~\bibnamefont {Iovino}}, \bibinfo {author} {\bibfnamefont
  {B.~R.}\ \bibnamefont {Granett}}, \bibinfo {author} {\bibfnamefont
  {U.}~\bibnamefont {Abbas}}, \bibinfo {author} {\bibfnamefont
  {C.}~\bibnamefont {Adami}}, \bibinfo {author} {\bibfnamefont
  {S.}~\bibnamefont {Arnouts}}, \bibinfo {author} {\bibfnamefont
  {J.}~\bibnamefont {Bel}}, \bibinfo {author} {\bibfnamefont {M.}~\bibnamefont
  {Bolzonella}}, \bibinfo {author} {\bibfnamefont {D.}~\bibnamefont {Bottini}},
  \bibinfo {author} {\bibfnamefont {A.}~\bibnamefont {Cappi}}, \bibinfo
  {author} {\bibfnamefont {J.}~\bibnamefont {Coupon}}, \bibinfo {author}
  {\bibfnamefont {O.}~\bibnamefont {Cucciati}}, \bibinfo {author}
  {\bibfnamefont {I.}~\bibnamefont {Davidzon}}, \bibinfo {author}
  {\bibfnamefont {G.}~\bibnamefont {De~Lucia}}, \bibinfo {author}
  {\bibfnamefont {A.}~\bibnamefont {Fritz}}, \bibinfo {author} {\bibfnamefont
  {P.}~\bibnamefont {Franzetti}}, \bibinfo {author} {\bibfnamefont
  {M.}~\bibnamefont {Fumana}}, \bibinfo {author} {\bibfnamefont
  {B.}~\bibnamefont {Garilli}}, \bibinfo {author} {\bibfnamefont
  {O.}~\bibnamefont {Ilbert}}, \bibinfo {author} {\bibfnamefont
  {J.}~\bibnamefont {Krywult}}, \bibinfo {author} {\bibfnamefont
  {V.}~\bibnamefont {Le~Brun}}, \bibinfo {author} {\bibfnamefont
  {O.}~\bibnamefont {Le~Fèvre}}, \bibinfo {author} {\bibfnamefont
  {D.}~\bibnamefont {Maccagni}}, \bibinfo {author} {\bibfnamefont
  {K.}~\bibnamefont {Małek}}, \bibinfo {author} {\bibfnamefont
  {F.}~\bibnamefont {Marulli}}, \bibinfo {author} {\bibfnamefont {H.~J.}\
  \bibnamefont {McCracken}}, \bibinfo {author} {\bibfnamefont {L.}~\bibnamefont
  {Moscardini}}, \bibinfo {author} {\bibfnamefont {L.}~\bibnamefont {Paioro}},
  \bibinfo {author} {\bibfnamefont {W.~J.}\ \bibnamefont {Percival}}, \bibinfo
  {author} {\bibfnamefont {M.}~\bibnamefont {Polletta}}, \bibinfo {author}
  {\bibfnamefont {A.}~\bibnamefont {Pollo}}, \bibinfo {author} {\bibfnamefont
  {H.}~\bibnamefont {Schlagenhaufer}}, \bibinfo {author} {\bibfnamefont
  {M.}~\bibnamefont {Scodeggio}}, \bibinfo {author} {\bibfnamefont {L.~A.~M.}\
  \bibnamefont {Tasca}}, \bibinfo {author} {\bibfnamefont {R.}~\bibnamefont
  {Tojeiro}}, \bibinfo {author} {\bibfnamefont {D.}~\bibnamefont {Vergani}},
  \bibinfo {author} {\bibfnamefont {A.}~\bibnamefont {Zanichelli}}, \bibinfo
  {author} {\bibfnamefont {A.}~\bibnamefont {Burden}}, \bibinfo {author}
  {\bibfnamefont {C.}~\bibnamefont {Di~Porto}}, \bibinfo {author}
  {\bibfnamefont {A.}~\bibnamefont {Marchetti}}, \bibinfo {author}
  {\bibfnamefont {C.}~\bibnamefont {Marinoni}}, \bibinfo {author}
  {\bibfnamefont {Y.}~\bibnamefont {Mellier}}, \bibinfo {author} {\bibfnamefont
  {P.}~\bibnamefont {Monaco}}, \bibinfo {author} {\bibfnamefont {R.~C.}\
  \bibnamefont {Nichol}}, \bibinfo {author} {\bibfnamefont {S.}~\bibnamefont
  {Phleps}}, \bibinfo {author} {\bibfnamefont {M.}~\bibnamefont {Wolk}}, \ and\
  \bibinfo {author} {\bibfnamefont {G.}~\bibnamefont {Zamorani}},\ }\href
  {\doibase 10.1051/0004-6361/201321463} {\bibfield  {journal} {\bibinfo
  {journal} {Astronomy and Astrophysics}\ }\textbf {\bibinfo {volume} {557}},\
  \bibinfo {pages} {A54} (\bibinfo {year} {2013})}\BibitemShut {NoStop}%
\bibitem [{\citenamefont {Howlett}\ \emph {et~al.}(2015)\citenamefont
  {Howlett}, \citenamefont {Ross}, \citenamefont {Samushia}, \citenamefont
  {Percival},\ and\ \citenamefont {Manera}}]{Howlett_2015}%
  \BibitemOpen
  \bibfield  {author} {\bibinfo {author} {\bibfnamefont {C.}~\bibnamefont
  {Howlett}}, \bibinfo {author} {\bibfnamefont {A.~J.}\ \bibnamefont {Ross}},
  \bibinfo {author} {\bibfnamefont {L.}~\bibnamefont {Samushia}}, \bibinfo
  {author} {\bibfnamefont {W.~J.}\ \bibnamefont {Percival}}, \ and\ \bibinfo
  {author} {\bibfnamefont {M.}~\bibnamefont {Manera}},\ }\href {\doibase
  10.1093/mnras/stu2693} {\bibfield  {journal} {\bibinfo  {journal} {Monthly
  Notices of the Royal Astronomical Society}\ }\textbf {\bibinfo {volume}
  {449}},\ \bibinfo {pages} {848–866} (\bibinfo {year} {2015})}\BibitemShut
  {NoStop}%
\bibitem [{\citenamefont {Feix}\ \emph {et~al.}(2015)\citenamefont {Feix},
  \citenamefont {Nusser},\ and\ \citenamefont {Branchini}}]{Feix:2015dla}%
  \BibitemOpen
  \bibfield  {author} {\bibinfo {author} {\bibfnamefont {M.}~\bibnamefont
  {Feix}}, \bibinfo {author} {\bibfnamefont {A.}~\bibnamefont {Nusser}}, \ and\
  \bibinfo {author} {\bibfnamefont {E.}~\bibnamefont {Branchini}},\ }\href
  {\doibase 10.1103/PhysRevLett.115.011301} {\bibfield  {journal} {\bibinfo
  {journal} {Phys. Rev. Lett.}\ }\textbf {\bibinfo {volume} {115}},\ \bibinfo
  {pages} {011301} (\bibinfo {year} {2015})},\ \Eprint
  {http://arxiv.org/abs/1503.05945} {arXiv:1503.05945 [astro-ph.CO]}
  \BibitemShut {NoStop}%
\bibitem [{\citenamefont {Okumura}\ \emph {et~al.}(2016)\citenamefont {Okumura}
  \emph {et~al.}}]{Okumura:2015lvp}%
  \BibitemOpen
  \bibfield  {author} {\bibinfo {author} {\bibfnamefont {T.}~\bibnamefont
  {Okumura}} \emph {et~al.},\ }\href {\doibase 10.1093/pasj/psw029} {\bibfield
  {journal} {\bibinfo  {journal} {Publ. Astron. Soc. Jap.}\ }\textbf {\bibinfo
  {volume} {68}},\ \bibinfo {pages} {38} (\bibinfo {year} {2016})},\ \Eprint
  {http://arxiv.org/abs/1511.08083} {arXiv:1511.08083 [astro-ph.CO]}
  \BibitemShut {NoStop}%
\bibitem [{\citenamefont {Huterer}\ \emph {et~al.}(2017)\citenamefont
  {Huterer}, \citenamefont {Shafer}, \citenamefont {Scolnic},\ and\
  \citenamefont {Schmidt}}]{Huterer:2016uyq}%
  \BibitemOpen
  \bibfield  {author} {\bibinfo {author} {\bibfnamefont {D.}~\bibnamefont
  {Huterer}}, \bibinfo {author} {\bibfnamefont {D.}~\bibnamefont {Shafer}},
  \bibinfo {author} {\bibfnamefont {D.}~\bibnamefont {Scolnic}}, \ and\
  \bibinfo {author} {\bibfnamefont {F.}~\bibnamefont {Schmidt}},\ }\href
  {\doibase 10.1088/1475-7516/2017/05/015} {\bibfield  {journal} {\bibinfo
  {journal} {JCAP}\ }\textbf {\bibinfo {volume} {05}},\ \bibinfo {pages} {015}
  (\bibinfo {year} {2017})},\ \Eprint {http://arxiv.org/abs/1611.09862}
  {arXiv:1611.09862 [astro-ph.CO]} \BibitemShut {NoStop}%
\bibitem [{\citenamefont {Pezzotta}\ \emph {et~al.}(2017)\citenamefont
  {Pezzotta} \emph {et~al.}}]{Pezzotta:2016gbo}%
  \BibitemOpen
  \bibfield  {author} {\bibinfo {author} {\bibfnamefont {A.}~\bibnamefont
  {Pezzotta}} \emph {et~al.},\ }\href {\doibase 10.1051/0004-6361/201630295}
  {\bibfield  {journal} {\bibinfo  {journal} {Astron. Astrophys.}\ }\textbf
  {\bibinfo {volume} {604}},\ \bibinfo {pages} {A33} (\bibinfo {year}
  {2017})},\ \Eprint {http://arxiv.org/abs/1612.05645} {arXiv:1612.05645
  [astro-ph.CO]} \BibitemShut {NoStop}%
\bibitem [{\citenamefont {Alam}\ \emph {et~al.}(2017)\citenamefont {Alam} \emph
  {et~al.}}]{BOSS:2016wmc}%
  \BibitemOpen
  \bibfield  {author} {\bibinfo {author} {\bibfnamefont {S.}~\bibnamefont
  {Alam}} \emph {et~al.} (\bibinfo {collaboration} {BOSS}),\ }\href {\doibase
  10.1093/mnras/stx721} {\bibfield  {journal} {\bibinfo  {journal} {Mon. Not.
  Roy. Astron. Soc.}\ }\textbf {\bibinfo {volume} {470}},\ \bibinfo {pages}
  {2617} (\bibinfo {year} {2017})},\ \Eprint {http://arxiv.org/abs/1607.03155}
  {arXiv:1607.03155 [astro-ph.CO]} \BibitemShut {NoStop}%
\bibitem [{\citenamefont {Nesseris}\ \emph {et~al.}(2017)\citenamefont
  {Nesseris}, \citenamefont {Pantazis},\ and\ \citenamefont
  {Perivolaropoulos}}]{Nesseris:2017vor}%
  \BibitemOpen
  \bibfield  {author} {\bibinfo {author} {\bibfnamefont {S.}~\bibnamefont
  {Nesseris}}, \bibinfo {author} {\bibfnamefont {G.}~\bibnamefont {Pantazis}},
  \ and\ \bibinfo {author} {\bibfnamefont {L.}~\bibnamefont
  {Perivolaropoulos}},\ }\href {\doibase 10.1103/PhysRevD.96.023542} {\bibfield
   {journal} {\bibinfo  {journal} {Phys. Rev. D}\ }\textbf {\bibinfo {volume}
  {96}},\ \bibinfo {pages} {023542} (\bibinfo {year} {2017})},\ \Eprint
  {http://arxiv.org/abs/1703.10538} {arXiv:1703.10538 [astro-ph.CO]}
  \BibitemShut {NoStop}%
\bibitem [{\citenamefont {Khatri}\ \emph {et~al.}(2024)\citenamefont {Khatri},
  \citenamefont {Singh},\ and\ \citenamefont {Srivastava}}]{Khatri:2024yfr}%
  \BibitemOpen
  \bibfield  {author} {\bibinfo {author} {\bibfnamefont {V.}~\bibnamefont
  {Khatri}}, \bibinfo {author} {\bibfnamefont {C.~P.}\ \bibnamefont {Singh}}, \
  and\ \bibinfo {author} {\bibfnamefont {M.}~\bibnamefont {Srivastava}},\
  }\href@noop {} {\  (\bibinfo {year} {2024})},\ \Eprint
  {http://arxiv.org/abs/2405.15296} {arXiv:2405.15296 [astro-ph.CO]}
  \BibitemShut {NoStop}%
\bibitem [{\citenamefont {da~Silva}\ and\ \citenamefont
  {Silva}(2021)}]{daSilva:2020mvk}%
  \BibitemOpen
  \bibfield  {author} {\bibinfo {author} {\bibfnamefont {W.~J.~C.}\
  \bibnamefont {da~Silva}}\ and\ \bibinfo {author} {\bibfnamefont
  {R.}~\bibnamefont {Silva}},\ }\href {\doibase
  10.1140/epjc/s10052-021-09177-7} {\bibfield  {journal} {\bibinfo  {journal}
  {Eur. Phys. J. C}\ }\textbf {\bibinfo {volume} {81}},\ \bibinfo {pages} {403}
  (\bibinfo {year} {2021})},\ \Eprint {http://arxiv.org/abs/2011.09516}
  {arXiv:2011.09516 [astro-ph.CO]} \BibitemShut {NoStop}%
\bibitem [{\citenamefont {Sahlu}\ \emph {et~al.}(2023)\citenamefont {Sahlu},
  \citenamefont {Mukhopadhyay}, \citenamefont {Mekuria},\ and\ \citenamefont
  {Abebe}}]{Sahlu:2023wvl}%
  \BibitemOpen
  \bibfield  {author} {\bibinfo {author} {\bibfnamefont {S.}~\bibnamefont
  {Sahlu}}, \bibinfo {author} {\bibfnamefont {U.}~\bibnamefont {Mukhopadhyay}},
  \bibinfo {author} {\bibfnamefont {R.~R.}\ \bibnamefont {Mekuria}}, \ and\
  \bibinfo {author} {\bibfnamefont {A.}~\bibnamefont {Abebe}},\ }\href@noop {}
  {\  (\bibinfo {year} {2023})},\ \Eprint {http://arxiv.org/abs/2301.02913}
  {arXiv:2301.02913 [astro-ph.CO]} \BibitemShut {NoStop}%
\bibitem [{\citenamefont {Kazantzidis}\ and\ \citenamefont
  {Perivolaropoulos}(2018)}]{Kazantzidis:2018rnb}%
  \BibitemOpen
  \bibfield  {author} {\bibinfo {author} {\bibfnamefont {L.}~\bibnamefont
  {Kazantzidis}}\ and\ \bibinfo {author} {\bibfnamefont {L.}~\bibnamefont
  {Perivolaropoulos}},\ }\href {\doibase 10.1103/PhysRevD.97.103503} {\bibfield
   {journal} {\bibinfo  {journal} {Phys. Rev. D}\ }\textbf {\bibinfo {volume}
  {97}},\ \bibinfo {pages} {103503} (\bibinfo {year} {2018})},\ \Eprint
  {http://arxiv.org/abs/1803.01337} {arXiv:1803.01337 [astro-ph.CO]}
  \BibitemShut {NoStop}%
\bibitem [{\citenamefont {Blake}\ \emph
  {et~al.}(2012{\natexlab{b}})\citenamefont {Blake}, \citenamefont {Brough},
  \citenamefont {Colless}, \citenamefont {Contreras}, \citenamefont {Couch},
  \citenamefont {Croom}, \citenamefont {Croton}, \citenamefont {Davis},
  \citenamefont {Drinkwater}, \citenamefont {Forster}, \citenamefont {Gilbank},
  \citenamefont {Gladders}, \citenamefont {Glazebrook}, \citenamefont
  {Jelliffe}, \citenamefont {Jurek}, \citenamefont {Li}, \citenamefont
  {Madore}, \citenamefont {Martin}, \citenamefont {Pimbblet}, \citenamefont
  {Poole}, \citenamefont {Pracy}, \citenamefont {Sharp}, \citenamefont
  {Wisnioski}, \citenamefont {Woods}, \citenamefont {Wyder},\ and\
  \citenamefont {Yee}}]{Blake2012TheWD}%
  \BibitemOpen
  \bibfield  {author} {\bibinfo {author} {\bibfnamefont {C.}~\bibnamefont
  {Blake}}, \bibinfo {author} {\bibfnamefont {S.}~\bibnamefont {Brough}},
  \bibinfo {author} {\bibfnamefont {M.}~\bibnamefont {Colless}}, \bibinfo
  {author} {\bibfnamefont {C.}~\bibnamefont {Contreras}}, \bibinfo {author}
  {\bibfnamefont {W.~J.}\ \bibnamefont {Couch}}, \bibinfo {author}
  {\bibfnamefont {S.~M.}\ \bibnamefont {Croom}}, \bibinfo {author}
  {\bibfnamefont {D.~J.}\ \bibnamefont {Croton}}, \bibinfo {author}
  {\bibfnamefont {T.~M.}\ \bibnamefont {Davis}}, \bibinfo {author}
  {\bibfnamefont {M.~J.}\ \bibnamefont {Drinkwater}}, \bibinfo {author}
  {\bibfnamefont {K.}~\bibnamefont {Forster}}, \bibinfo {author} {\bibfnamefont
  {D.}~\bibnamefont {Gilbank}}, \bibinfo {author} {\bibfnamefont {M.~D.}\
  \bibnamefont {Gladders}}, \bibinfo {author} {\bibfnamefont {K.}~\bibnamefont
  {Glazebrook}}, \bibinfo {author} {\bibfnamefont {B.}~\bibnamefont
  {Jelliffe}}, \bibinfo {author} {\bibfnamefont {R.~J.}\ \bibnamefont {Jurek}},
  \bibinfo {author} {\bibfnamefont {I.~H.}\ \bibnamefont {Li}}, \bibinfo
  {author} {\bibfnamefont {B.~F.}\ \bibnamefont {Madore}}, \bibinfo {author}
  {\bibfnamefont {C.~D.}\ \bibnamefont {Martin}}, \bibinfo {author}
  {\bibfnamefont {K.~A.}\ \bibnamefont {Pimbblet}}, \bibinfo {author}
  {\bibfnamefont {G.~B.}\ \bibnamefont {Poole}}, \bibinfo {author}
  {\bibfnamefont {M.~B.}\ \bibnamefont {Pracy}}, \bibinfo {author}
  {\bibfnamefont {R.}~\bibnamefont {Sharp}}, \bibinfo {author} {\bibfnamefont
  {E.}~\bibnamefont {Wisnioski}}, \bibinfo {author} {\bibfnamefont
  {D.}~\bibnamefont {Woods}}, \bibinfo {author} {\bibfnamefont {T.~K.}\
  \bibnamefont {Wyder}}, \ and\ \bibinfo {author} {\bibfnamefont {H.~K.~C.}\
  \bibnamefont {Yee}},\ }\href
  {https://api.semanticscholar.org/CorpusID:28284545} {\bibfield  {journal}
  {\bibinfo  {journal} {Monthly Notices of the Royal Astronomical Society}\
  }\textbf {\bibinfo {volume} {425}},\ \bibinfo {pages} {405} (\bibinfo {year}
  {2012}{\natexlab{b}})}\BibitemShut {NoStop}%
\bibitem [{\citenamefont {Zhao}\ \emph {et~al.}(2018)\citenamefont {Zhao},
  \citenamefont {Wang}, \citenamefont {Saito}, \citenamefont {Gil-Marín},
  \citenamefont {Percival}, \citenamefont {Wang}, \citenamefont {Chuang},
  \citenamefont {Ruggeri}, \citenamefont {Mueller}, \citenamefont {Zhu},
  \citenamefont {Ross}, \citenamefont {Tojeiro}, \citenamefont {Pâris},
  \citenamefont {Myers}, \citenamefont {Tinker}, \citenamefont {Li},
  \citenamefont {Burtin}, \citenamefont {Zarrouk}, \citenamefont {Beutler},
  \citenamefont {Baumgarten}, \citenamefont {Bautista}, \citenamefont
  {Brownstein}, \citenamefont {Dawson}, \citenamefont {Hou}, \citenamefont
  {de la Macorra}, \citenamefont {Rossi}, \citenamefont {Peacock},
  \citenamefont {Sánchez}, \citenamefont {Shafieloo}, \citenamefont
  {Schneider},\ and\ \citenamefont {Zhao}}]{10.1093/mnras/sty2845}%
  \BibitemOpen
  \bibfield  {author} {\bibinfo {author} {\bibfnamefont {G.-B.}\ \bibnamefont
  {Zhao}}, \bibinfo {author} {\bibfnamefont {Y.}~\bibnamefont {Wang}}, \bibinfo
  {author} {\bibfnamefont {S.}~\bibnamefont {Saito}}, \bibinfo {author}
  {\bibfnamefont {H.}~\bibnamefont {Gil-Marín}}, \bibinfo {author}
  {\bibfnamefont {W.~J.}\ \bibnamefont {Percival}}, \bibinfo {author}
  {\bibfnamefont {D.}~\bibnamefont {Wang}}, \bibinfo {author} {\bibfnamefont
  {C.-H.}\ \bibnamefont {Chuang}}, \bibinfo {author} {\bibfnamefont
  {R.}~\bibnamefont {Ruggeri}}, \bibinfo {author} {\bibfnamefont {E.-M.}\
  \bibnamefont {Mueller}}, \bibinfo {author} {\bibfnamefont {F.}~\bibnamefont
  {Zhu}}, \bibinfo {author} {\bibfnamefont {A.~J.}\ \bibnamefont {Ross}},
  \bibinfo {author} {\bibfnamefont {R.}~\bibnamefont {Tojeiro}}, \bibinfo
  {author} {\bibfnamefont {I.}~\bibnamefont {Pâris}}, \bibinfo {author}
  {\bibfnamefont {A.~D.}\ \bibnamefont {Myers}}, \bibinfo {author}
  {\bibfnamefont {J.~L.}\ \bibnamefont {Tinker}}, \bibinfo {author}
  {\bibfnamefont {J.}~\bibnamefont {Li}}, \bibinfo {author} {\bibfnamefont
  {E.}~\bibnamefont {Burtin}}, \bibinfo {author} {\bibfnamefont
  {P.}~\bibnamefont {Zarrouk}}, \bibinfo {author} {\bibfnamefont
  {F.}~\bibnamefont {Beutler}}, \bibinfo {author} {\bibfnamefont
  {F.}~\bibnamefont {Baumgarten}}, \bibinfo {author} {\bibfnamefont {J.~E.}\
  \bibnamefont {Bautista}}, \bibinfo {author} {\bibfnamefont {J.~R.}\
  \bibnamefont {Brownstein}}, \bibinfo {author} {\bibfnamefont {K.~S.}\
  \bibnamefont {Dawson}}, \bibinfo {author} {\bibfnamefont {J.}~\bibnamefont
  {Hou}}, \bibinfo {author} {\bibfnamefont {A.}~\bibnamefont
  {de la Macorra}}, \bibinfo {author} {\bibfnamefont {G.}~\bibnamefont
  {Rossi}}, \bibinfo {author} {\bibfnamefont {J.~A.}\ \bibnamefont {Peacock}},
  \bibinfo {author} {\bibfnamefont {A.~G.}\ \bibnamefont {Sánchez}}, \bibinfo
  {author} {\bibfnamefont {A.}~\bibnamefont {Shafieloo}}, \bibinfo {author}
  {\bibfnamefont {D.~P.}\ \bibnamefont {Schneider}}, \ and\ \bibinfo {author}
  {\bibfnamefont {C.}~\bibnamefont {Zhao}},\ }\href {\doibase
  10.1093/mnras/sty2845} {\bibfield  {journal} {\bibinfo  {journal} {Monthly
  Notices of the Royal Astronomical Society}\ }\textbf {\bibinfo {volume}
  {482}},\ \bibinfo {pages} {3497} (\bibinfo {year} {2018})}\BibitemShut
  {NoStop}%
\bibitem [{\citenamefont {Foreman-Mackey}\ \emph {et~al.}(2013)\citenamefont
  {Foreman-Mackey}, \citenamefont {Hogg}, \citenamefont {Lang},\ and\
  \citenamefont {Goodman}}]{emcee}%
  \BibitemOpen
  \bibfield  {author} {\bibinfo {author} {\bibfnamefont {D.}~\bibnamefont
  {Foreman-Mackey}}, \bibinfo {author} {\bibfnamefont {D.~W.}\ \bibnamefont
  {Hogg}}, \bibinfo {author} {\bibfnamefont {D.}~\bibnamefont {Lang}}, \ and\
  \bibinfo {author} {\bibfnamefont {J.}~\bibnamefont {Goodman}},\ }\href
  {\doibase 10.1086/670067} {\bibfield  {journal} {\bibinfo  {journal}
  {Publications of the Astronomical Society of the Pacific}\ }\textbf {\bibinfo
  {volume} {125}},\ \bibinfo {pages} {306–312} (\bibinfo {year}
  {2013})}\BibitemShut {NoStop}%
\bibitem [{\citenamefont {Heymans}\ \emph {et~al.}(2021)\citenamefont {Heymans}
  \emph {et~al.}}]{Heymans:2020gsg}%
  \BibitemOpen
  \bibfield  {author} {\bibinfo {author} {\bibfnamefont {C.}~\bibnamefont
  {Heymans}} \emph {et~al.},\ }\href {\doibase 10.1051/0004-6361/202039063}
  {\bibfield  {journal} {\bibinfo  {journal} {Astron. Astrophys.}\ }\textbf
  {\bibinfo {volume} {646}},\ \bibinfo {pages} {A140} (\bibinfo {year}
  {2021})},\ \Eprint {http://arxiv.org/abs/2007.15632} {arXiv:2007.15632
  [astro-ph.CO]} \BibitemShut {NoStop}%
\bibitem [{\citenamefont {Lewis}(2019)}]{getdist}%
  \BibitemOpen
  \bibfield  {author} {\bibinfo {author} {\bibfnamefont {A.}~\bibnamefont
  {Lewis}},\ }\href {https://getdist.readthedocs.io} {\  (\bibinfo {year}
  {2019})},\ \Eprint {http://arxiv.org/abs/1910.13970} {arXiv:1910.13970
  [astro-ph.IM]} \BibitemShut {NoStop}%
\bibitem [{\citenamefont {Akaike}\ and\ \citenamefont
  {H.}(1974)}]{Akaike1974A}%
  \BibitemOpen
  \bibfield  {author} {\bibinfo {author} {\bibnamefont {Akaike}}\ and\ \bibinfo
  {author} {\bibnamefont {H.}},\ }\href@noop {} {\bibfield  {journal} {\bibinfo
   {journal} {Automatic Control, IEEE Transactions on}\ } (\bibinfo {year}
  {1974})}\BibitemShut {NoStop}%
\bibitem [{\citenamefont {{Schwarz}}(1978)}]{1978AnSta...6..461S}%
  \BibitemOpen
  \bibfield  {author} {\bibinfo {author} {\bibfnamefont {G.}~\bibnamefont
  {{Schwarz}}},\ }\href@noop {} {\bibfield  {journal} {\bibinfo  {journal}
  {Annals of Statistics}\ }\textbf {\bibinfo {volume} {6}},\ \bibinfo {pages}
  {461} (\bibinfo {year} {1978})}\BibitemShut {NoStop}%
\bibitem [{\citenamefont {Kass}\ and\ \citenamefont
  {and}(1995)}]{Kass01061995}%
  \BibitemOpen
  \bibfield  {author} {\bibinfo {author} {\bibfnamefont {R.~E.}\ \bibnamefont
  {Kass}}\ and\ \bibinfo {author} {\bibfnamefont {A.~E.~R.}\ \bibnamefont
  {and}},\ }\href {\doibase 10.1080/01621459.1995.10476572} {\bibfield
  {journal} {\bibinfo  {journal} {Journal of the American Statistical
  Association}\ }\textbf {\bibinfo {volume} {90}},\ \bibinfo {pages} {773}
  (\bibinfo {year} {1995})}\BibitemShut {NoStop}%
\bibitem [{\citenamefont {Wagoner}\ \emph {et~al.}(2021)\citenamefont
  {Wagoner}, \citenamefont {Rozo}, \citenamefont {Aung},\ and\ \citenamefont
  {Nagai}}]{Wagoner:2020wht}%
  \BibitemOpen
  \bibfield  {author} {\bibinfo {author} {\bibfnamefont {E.~L.}\ \bibnamefont
  {Wagoner}}, \bibinfo {author} {\bibfnamefont {E.}~\bibnamefont {Rozo}},
  \bibinfo {author} {\bibfnamefont {H.}~\bibnamefont {Aung}}, \ and\ \bibinfo
  {author} {\bibfnamefont {D.}~\bibnamefont {Nagai}},\ }\href {\doibase
  10.1093/mnras/stab1012} {\bibfield  {journal} {\bibinfo  {journal} {Mon. Not.
  Roy. Astron. Soc.}\ }\textbf {\bibinfo {volume} {504}},\ \bibinfo {pages}
  {1619} (\bibinfo {year} {2021})},\ \Eprint {http://arxiv.org/abs/2010.11324}
  {arXiv:2010.11324 [astro-ph.CO]} \BibitemShut {NoStop}%
\bibitem [{\citenamefont {Wang}(2021)}]{Wang:2020hqq}%
  \BibitemOpen
  \bibfield  {author} {\bibinfo {author} {\bibfnamefont {D.}~\bibnamefont
  {Wang}},\ }\href {\doibase 10.1103/PhysRevD.103.123538} {\bibfield  {journal}
  {\bibinfo  {journal} {Phys. Rev. D}\ }\textbf {\bibinfo {volume} {103}},\
  \bibinfo {pages} {123538} (\bibinfo {year} {2021})},\ \Eprint
  {http://arxiv.org/abs/2011.11924} {arXiv:2011.11924 [astro-ph.CO]}
  \BibitemShut {NoStop}%
\bibitem [{\citenamefont {Wang}(2018)}]{Wang:2018ahw}%
  \BibitemOpen
  \bibfield  {author} {\bibinfo {author} {\bibfnamefont {D.}~\bibnamefont
  {Wang}},\ }\href {\doibase 10.1103/PhysRevD.97.123507} {\bibfield  {journal}
  {\bibinfo  {journal} {Phys. Rev. D}\ }\textbf {\bibinfo {volume} {97}},\
  \bibinfo {pages} {123507} (\bibinfo {year} {2018})},\ \Eprint
  {http://arxiv.org/abs/1801.02371} {arXiv:1801.02371 [astro-ph.CO]}
  \BibitemShut {NoStop}%
\bibitem [{\citenamefont {Wang}\ and\ \citenamefont
  {Meng}(2017{\natexlab{a}})}]{Wang:2017fcr}%
  \BibitemOpen
  \bibfield  {author} {\bibinfo {author} {\bibfnamefont {D.}~\bibnamefont
  {Wang}}\ and\ \bibinfo {author} {\bibfnamefont {X.-H.}\ \bibnamefont
  {Meng}},\ }\href {\doibase 10.1016/j.dark.2017.09.005} {\bibfield  {journal}
  {\bibinfo  {journal} {Phys. Dark Univ.}\ }\textbf {\bibinfo {volume} {18}},\
  \bibinfo {pages} {30} (\bibinfo {year} {2017}{\natexlab{a}})},\ \Eprint
  {http://arxiv.org/abs/1709.04134} {arXiv:1709.04134 [astro-ph.CO]}
  \BibitemShut {NoStop}%
\bibitem [{\citenamefont {Wang}\ and\ \citenamefont
  {Meng}(2017{\natexlab{b}})}]{Wang:2016pag}%
  \BibitemOpen
  \bibfield  {author} {\bibinfo {author} {\bibfnamefont {D.}~\bibnamefont
  {Wang}}\ and\ \bibinfo {author} {\bibfnamefont {X.-H.}\ \bibnamefont
  {Meng}},\ }\href {\doibase 10.3847/1538-4357/aa667e} {\bibfield  {journal}
  {\bibinfo  {journal} {Astrophys. J.}\ }\textbf {\bibinfo {volume} {843}},\
  \bibinfo {pages} {100} (\bibinfo {year} {2017}{\natexlab{b}})},\ \Eprint
  {http://arxiv.org/abs/1612.09023} {arXiv:1612.09023 [astro-ph.CO]}
  \BibitemShut {NoStop}%
\bibitem [{\citenamefont {Li}\ \emph {et~al.}(2024)\citenamefont {Li},
  \citenamefont {Wu}, \citenamefont {Du}, \citenamefont {Jin}, \citenamefont
  {Li}, \citenamefont {Zhang},\ and\ \citenamefont {Zhang}}]{Li:2024qso}%
  \BibitemOpen
  \bibfield  {author} {\bibinfo {author} {\bibfnamefont {T.-N.}\ \bibnamefont
  {Li}}, \bibinfo {author} {\bibfnamefont {P.-J.}\ \bibnamefont {Wu}}, \bibinfo
  {author} {\bibfnamefont {G.-H.}\ \bibnamefont {Du}}, \bibinfo {author}
  {\bibfnamefont {S.-J.}\ \bibnamefont {Jin}}, \bibinfo {author} {\bibfnamefont
  {H.-L.}\ \bibnamefont {Li}}, \bibinfo {author} {\bibfnamefont {J.-F.}\
  \bibnamefont {Zhang}}, \ and\ \bibinfo {author} {\bibfnamefont
  {X.}~\bibnamefont {Zhang}},\ }\href {\doibase 10.3847/1538-4357/ad87f0}
  {\bibfield  {journal} {\bibinfo  {journal} {Astrophys. J.}\ }\textbf
  {\bibinfo {volume} {976}},\ \bibinfo {pages} {1} (\bibinfo {year} {2024})},\
  \Eprint {http://arxiv.org/abs/2407.14934} {arXiv:2407.14934 [astro-ph.CO]}
  \BibitemShut {NoStop}%
\bibitem [{\citenamefont {Pan}\ \emph {et~al.}(2025)\citenamefont {Pan},
  \citenamefont {Paul}, \citenamefont {Saridakis},\ and\ \citenamefont
  {Yang}}]{Pan:2025qwy}%
  \BibitemOpen
  \bibfield  {author} {\bibinfo {author} {\bibfnamefont {S.}~\bibnamefont
  {Pan}}, \bibinfo {author} {\bibfnamefont {S.}~\bibnamefont {Paul}}, \bibinfo
  {author} {\bibfnamefont {E.~N.}\ \bibnamefont {Saridakis}}, \ and\ \bibinfo
  {author} {\bibfnamefont {W.}~\bibnamefont {Yang}},\ }\href@noop {} {\
  (\bibinfo {year} {2025})},\ \Eprint {http://arxiv.org/abs/2504.00994}
  {arXiv:2504.00994 [astro-ph.CO]} \BibitemShut {NoStop}%
\bibitem [{\citenamefont {Abdul~Karim}\ \emph {et~al.}(2025)\citenamefont
  {Abdul~Karim} \emph {et~al.}}]{DESI:2025zgx}%
  \BibitemOpen
  \bibfield  {author} {\bibinfo {author} {\bibfnamefont {M.}~\bibnamefont
  {Abdul~Karim}} \emph {et~al.} (\bibinfo {collaboration} {DESI}),\ }\href@noop
  {} {\  (\bibinfo {year} {2025})},\ \Eprint {http://arxiv.org/abs/2503.14738}
  {arXiv:2503.14738 [astro-ph.CO]} \BibitemShut {NoStop}%
\bibitem [{\citenamefont {Sol\`a~Peracaula}\ \emph {et~al.}(2021)\citenamefont
  {Sol\`a~Peracaula}, \citenamefont {G\'omez-Valent}, \citenamefont
  {de~Cruz~Perez},\ and\ \citenamefont
  {Moreno-Pulido}}]{SolaPeracaula:2021gxi}%
  \BibitemOpen
  \bibfield  {author} {\bibinfo {author} {\bibfnamefont {J.}~\bibnamefont
  {Sol\`a~Peracaula}}, \bibinfo {author} {\bibfnamefont {A.}~\bibnamefont
  {G\'omez-Valent}}, \bibinfo {author} {\bibfnamefont {J.}~\bibnamefont
  {de~Cruz~Perez}}, \ and\ \bibinfo {author} {\bibfnamefont {C.}~\bibnamefont
  {Moreno-Pulido}},\ }\href {\doibase 10.1209/0295-5075/134/19001} {\bibfield
  {journal} {\bibinfo  {journal} {EPL}\ }\textbf {\bibinfo {volume} {134}},\
  \bibinfo {pages} {19001} (\bibinfo {year} {2021})},\ \Eprint
  {http://arxiv.org/abs/2102.12758} {arXiv:2102.12758 [astro-ph.CO]}
  \BibitemShut {NoStop}%
\bibitem [{\citenamefont {Eisenstein}\ and\ \citenamefont
  {Hu}(1998)}]{Eisenstein:1997ik}%
  \BibitemOpen
  \bibfield  {author} {\bibinfo {author} {\bibfnamefont {D.~J.}\ \bibnamefont
  {Eisenstein}}\ and\ \bibinfo {author} {\bibfnamefont {W.}~\bibnamefont
  {Hu}},\ }\href {\doibase 10.1086/305424} {\bibfield  {journal} {\bibinfo
  {journal} {Astrophys. J.}\ }\textbf {\bibinfo {volume} {496}},\ \bibinfo
  {pages} {605} (\bibinfo {year} {1998})},\ \Eprint
  {http://arxiv.org/abs/astro-ph/9709112} {arXiv:astro-ph/9709112} \BibitemShut
  {NoStop}%
\end{thebibliography}%

\end{document}